\documentclass[12pt,preprint]{aastex}
\newcommand{\mdot}{\dot M }
\newcommand{\va}{v_{A} }
\newcommand{\bs}{B_0 }
\newcommand{\rs}{r_0 }
\newcommand{\ra}{r_{A} }
\newcommand{\vp}{{\mathcal{V}}_0 }

\shorttitle{Angular Momemtum Evolution in Low-Mass Stars.  I.}
\shortauthors{Reiners \& Mohanty}

\begin{document}


\title{Radius Dependent Angular Momentum Evolution\\ in Low-Mass
  Stars.  I.}

% II. Rotation, Magnetism, and Chromospheric Activity}


\author{Ansgar Reiners}
\affil{Institut f\"ur Astrophysik, Georg-August-Universit\"at,
  Friedrich-Hund-Platz 1, 37077 G\"ottingen, Germany}
\email{Ansgar.Reiners@phys.uni-goettingen.de}
\and
\author{Subhanjoy Mohanty}
\affil{Imperial College London, 1010 Blackett Lab., Prince Consort Road, London SW7 2AZ, UK}
%\email{basri@berkeley.edu}


%\altaffiltext{*}{Emmy Noether Fellow}


\begin{abstract}
  Angular momentum evolution in low-mass stars is determined by
  initial conditions during star formation, stellar structure
  evolution, and the behaviour of stellar magnetic fields.  Here we
  show that the empirical picture of angular momentum evolution arises
  naturally if rotation is related to magnetic field strength instead
  of to magnetic flux, and formulate a corrected braking law based on
  this.  Angular momentum evolution then becomes a strong function of
  stellar radius, explaining the main trends observed in open clusters
  and field stars at a few Gyr: the steep transition in rotation at
  the boundary to full convection arises primarily from the large
  change in radius across this boundary, and does not require changes
  in dynamo mode or field topology.  Additionally, the data suggest
  transient core-envelope decoupling among solar-type stars, and field
  saturation at longer periods in very low-mass stars.  For solar-type
  stars, our model is also in good agreement with the empirical
  Skumanich law.  Finally, in further support of the theory, we show
  that the predicted age at which low-mass stars spin down from the
  saturated to unsaturated field regimes in our model corresponds
  remarkably well to the observed lifetime of magnetic activity in
  these stars.
\end{abstract}

\section{Introduction}

Angular momentum evolution in low-mass stars is the result of a
complex interplay between initial conditions during star formation,
the evolution of stellar structure, and the behaviour of stellar winds
and magnetic fields. Data on stellar rotation at different ages thus
provide invaluable insights into star-formation and evolution.  More
than four decades ago, \citet{1970saac.book..385K} discussed the
distribution of angular momentum ($J$) with stellar mass and showed
that in stars more massive than $\sim$ 1.5$\,M_\odot$, $J$ is
proportional to mass, consistent with conserved stellar angular
momentum since their formation. Lower-mass stars, in contrast, evince
much lower angular momenta, an effect ascribed to magnetic braking due
to coupling between an ionised wind and magnetic fields spawned by an
internal stellar dynamo \citep{1962AnAp...25...18S,
  1967ApJ...148..217W, 1968MNRAS.138..359M, 1984LNP...193...49M} (this
does not occur in higher mass stars, presumably because dynamoes
cannot exist in their mainly radiative interiors).

Specifically, during Pre-Main Sequence (PMS) evolution, the rotation
of a low-mass star is regulated by the competition between spin-up due
to contraction at constant angular momentum, and spin-down due to
magnetic braking by a wind that actively removes angular momentum.
Contraction ceases once the star arrives on the Main Sequence (MS),
and wind-driven braking spins the star down for the rest of its MS
lifetime.

A large amount of rotation data for low-mass stars has been gathered
over the past decades, both in the field and in clusters of various
ages, in order to determine the precise rules underlying the
qualitative picture of angular momentum evolution painted above
\citep{2007ApJ...669.1167B, 2008ApJ...684.1390R, 2009IAUS..258..363I}.
These efforts are still hampered to some degree by observational bias:
for instance, in a normal ground-based observing campaign, it is
difficult to detect rotation periods shorter than a few hours or
longer than a few weeks.  This may skew our view of the distribution
of angular momenta at any age, and in particular of the distribution
at very early ages, which is currently required as an empirical
initial condition for models of rotational evolution.  Nevertheless, a
broad physical framework has emerged.  Faster rotation generates
stronger magnetic fields, and it is the field stresses that transfer
the stellar angular momentum to the outflowing winds, thereby braking
the star.  The rate of angular momentum loss is thus proportional to
the angular velocity of rotation raised to some power, where the
exponent depends on the magnetic field geometry
\citep{1984LNP...193...49M, 1988ApJ...333..236K}.  Moreover, the data
also require that this exponent change at some critical rotation
velocity (the ``saturation'' velocity), such that the loss rate
increases more slowly with rotation above this threshold
\citep{1995ApJ...441..865C, 2000ApJ...534..335S}. Models incorporating
these features, and using the observed angular momenta in very young
star-forming regions and in the present-day Sun as the initial and
final boundary conditions respectively, can reproduce the rotation
data for solar-mass stars at various ages with reasonable success.  An
additional effect which may play a role in solar-type stars is
rotational decoupling between the radiative core and the convective
envelope \citep{2008AA...489L..53B, 2009IAUS..258..363I}; we revisit
this point later.

What this theoretical picture is currently {\it not} able to do is
{\it simultaneously} match the rotational evolution of {\it very}
low-mass stars (VLMS; $\le 0.5$\,M$_{\odot}$)
\citep{2009IAUS..258..363I}.  This is our goal here.  We first
summarise the observed trends in the rotation of these stars, and then
describe our modelling.

\section{Empirical Framework}

A compilation of more than 3000 stellar rotation rates spread over
different ages is presented by \citet{2009IAUS..258..363I}. It reveals
that in very young clusters ($<10$\,Myr) the distribution of rotation
rates is relatively uniform, with periods between 1 and 10\,d for
stars over the entire mass range $\sim$ 1.5--0.1\,M$_{\odot}$. By ages
between 100\,Myr and 200\,Myr, though, the situation has changed.
Stars with mass $>$0.5\,M$_{\odot}$ mostly evince periods of a few to
10\,d, but the VLMS now show a clear trend of faster rotation with
decreasing mass: the maximum period drops from $\sim$10\,d at
0.5\,M$_{\odot}$ to $<$$\sim$ 1\,d by 0.2\,M$_{\odot}$.  At later ages
of around 600\,Myr, the data are more sparse but the same trend is
still evident: stars more massive than $\sim$0.5\,M$_{\odot}$ have
mostly slowed to periods around 10\,d, while in the VLMS the rotation
period declines rapidly with lower stellar mass.  Finally, the trend
is repeated in field stars at ages of a few Gyr (corresponding to the
young disk population).  Rotation rates are now generally very low for
stars above 0.4\,M$_{\odot}$: observed periods are of order that of
the Sun, 10--20\,d, and projected surface velocities ($v\,\sin{i}$)
below typical detection limits. Young disk VLMS below
0.4\,M$_{\odot}$, on the other hand, again manifest a steep dropoff in
rotation period with decreasing stellar mass, with periods down to a
fraction of a day\footnote{\citet{2011ApJ...727...56I} have recently
  found a population of extremely slowly rotating M dwarfs, with
  periods of $\sim$20--100\,d, but all these appear to be
  old-disk/halo stars at ages of $\sim$10 Gyr: much older than the
  young-disk field stars at a few Gyr we are referring to here.  We
  discuss the old population later in the paper.}.  The transition
between the two regimes is abrupt, and coincides with the boundary at
which stars become fully convective.

The theory of angular momentum evolution as currently formulated can
explain neither the mass-dependent periods nor the very fast rotation
in the field seen in the VLMS.  For example, the theory predicts all
VLMS to have converged to very slow rotation by a few Gyr, contrary to
observations \citep{2009IAUS..258..363I}.  Various prescriptive fixes
have been suggested, invoking a dependence of angular momentum
evolution on parameters ranging from mass- and rotation-dependent wind
velocities \citep{2009IAUS..258..363I, 2011ApJ...727...56I} to
convective overturn timescales \citep{2010ApJ...721..675B}.  None of
these fixes provide a theoretical motivation, however, for the
particular form of the dependencies invoked.  For instance,
\citet{2000ApJ...534..335S}, using a standard Rossby number scaling
for the saturation velocity, find that ``neither solid-body rotation
nor differentially rotating models can simultaneously reproduce the
observed stellar spin-down in the 0.6--1.1 M$_{\odot}$ range and for
stars between 0.1 and 0.6 M$_{\odot}$''.  Consequently, they argue
that ``the saturation threshold drops more steeply at low masses than
would be predicted by a simple Rossby number scaling'', and are forced
to adopt an ad hoc scaling to match the very low mass data.  The fact
that the turnover in periods in field stars happens close to the
stellar mass where the interiors become fully convective has moreover
led to the suggestion that changes in the dynamo or field topology in
fully convective stars may drive the turnover
\citep{2009IAUS..258..363I}, but how this actually works is also
unspecified.  Here we reexamine the basic formulation of the theory of
angular momentum evolution for low-mass stars in general, and show
that a fundamental dependence on radius has previously been missed.
The error arises due to a confusion between magnetic field strength
and magnetic flux in the formulation by \citet[][ hereafter
K88]{1988ApJ...333..236K}, which has been widely adopted.  Accounting
for this effect allows all low-mass stars, from solar-mass to VLMS, to
be broadly fit by the same theory.

In \S\S 3--5, we present our new formulation of angular momentum
evolution, under the simplest possible physically motivated
assumptions.  After a brief discussion of the methods we use to
determine stellar masses (\S6), and of our adopted initial conditions
(\S7), we compare our model predictions to data for young clusters and
field stars in \S8.  We show that the overall empirical picture of
rotation evolution is well reproduced by our theory.  Concurrently, we
point out remaining discrepancies between the model and data, and
explain them via secondary physical effects not included in our theory
from the outset, namely: {\it (i)} transient core-envelope decoupling
in solar-type stars; and {\it (ii)} Rossby-number scaling of the
saturation velocity, particularly important for very old VLMS (old
disk / halo ages).  As further support for our model, we show that it
also predicts the observed lifetime of magnetic activity in low-mass
stars remarkably well.  Finally, in \S9 we discuss possible
sophistications (e.g., in the magnetic topology) that may be
introduced into the theory to reflect reality still better.

\section{Formalism for Angular Momentum Evolution}

Following the work of \citet{1968MNRAS.138..359M, 1983IAUS..102..449R}
and Roxburgh \citet{1983IAUS..102..449R}, we assume that the magnetic
field enforces corotation of the outflowing coronal gas out to some
radius $r_A$, at which point the wind velocity becomes Alfv\'enic, so
that the wind flows freely beyond $r_A$.  We further make the usual
simplifying assumption that the Alfv\'en surface $S_A$ is
approximately spherical, i.e., defined by a single radius $r_A$.
Finally, we assume that {\it the magnetic field is radial everywhere}
(i.e., of multipole order $m$ = 2).  Higher-order multipoles present
no problems for the formalism below
\citep[see][]{1983IAUS..102..449R}. However, we choose the simplest
field geometry here to show that the main observed features of angular
momentum evolution in low-mass stars do not require variations of the
field geometry with stellar mass, but arise out of basic
considerations of the relationship between the field strength and
rotation.  The limitations of a radial field, and improvements
possible by considering more realistic field topologies, are discussed
at the end in \S9.  Note that K88, and current models based on it, use
$m$ = 2.25, very close to our purely radial field $m$ = 2.

We denote stellar mass, radius, angular velocity and mass loss rate by
$M$, $R$, $\Omega$ and $\mdot$, and radial distance from the stellar
centre by $r$.  All values at the base of the wind are denoted by the
subscript `0'.  We adopt the usual approximation that the wind base is
essentially at the stellar surface, so $\rs$ $\approx$ $R$.  All
quantities on the Alfv\'en surface are denoted by the subscript `$A$'.
Following \citet[][hereafter M84]{1984LNP...193...49M}, we parametrize
the average magnetic field strength on the stellar surface, $\bs$, as
a power-law in the stellar angular velocity:
\begin{equation} 
  \label{eq:Bs}
  \bs \propto {\Omega}^a
\end{equation}
where $a$ likely varies between 1 and 2 for unsaturated fields (regime
where field strength increases with rotation) and drops to 0 by
definition when the field strength saturates (i.e., remains constant
with increasing rotation; saturation discussed further below).  Such a
dependence of field strength on angular velocity is expected for both
the $\alpha\Omega$ dynamo postulated for solar-type stars with a
radiative-convective interface \citep[][and references
therein]{1972Ap&SS..15..307D}, and the $\alpha^2$ dynamo that may
operate in fully convective VLMS \citep{2006A&A...446.1027C}. Note
here the critical difference between our parametrization and that by
K88, who assumes that the surface magnetic {\it flux} goes as some
power of the angular velocity ($\bs R^2 \propto \Omega^a$) instead of
the magnetic field {\it strength} obeying this relationship (equation
[1]).  The functional form we adopt is empirically observed
\citep{1996IAUS..176..237S}, where it is often written in the form $fB
\propto \Omega^a$.  Here $B$ is the strength of the {\it observed}
field and $f$ its areal covering fraction, making $fB$ the estimated
{\it mean} surface field strength, corresponding exactly to $\bs$ in
our notation.  While K88 refers to the latter empirical studies, he
erroneously concludes that $fB$ is the total flux, when it is really
the {\it observed} flux {\it normalised by the total stellar surface
  area}, i.e., the mean flux {\it density}, which is just the mean
field strength.  The error is preserved in many later works based on
the K88 formalism; the factor of $R^2$ it introduces has serious
repercussions.

Our formulation above implicitly demands defining the critical
(saturation) rotation velocity ${\Omega}_{\rm crit}$ where the switch
from the unsaturated to saturated regime occurs, and the (constant)
field strength $B_{\rm crit}$ in the saturated regime.  We thus
explicitly rewrite equation [1] for the field strength as:
\begin{eqnarray}
  \label{eq:Bscale}
 B_0 = B_{\rm crit} \,\,\,\,\,\,\,\,\,\, {\rm for} \,\,\,\,\, \Omega \ge \Omega_{\rm crit} \,\,\,\,\, {\rm (saturated)} \,\,\,\,\,\, \nonumber \\
 B_0 = B_{\rm crit}\left(\frac{\Omega}{\Omega_{\rm crit}}\right)^a \,\,\,\,\,\,\,\,\,\, {\rm for} \,\,\,\,\, \Omega < \Omega_{\rm crit} \,\,\,\,\, {\rm (unsaturated)}
\end{eqnarray}
This form of the field scaling is motivated by theory as well as
observations of sun-like stars and M-stars
\citep{2009ApJ...692..538R}.  Specifically, we adopt an exponent of $a
= 1.5$ in the unsaturated regime, within the range 1--2 expected from
theoretical considerations and consistent with the data (e.g.,
\citet{1996IAUS..176..237S} finds $fB \propto \Omega^{1.7}$).
$\Omega_{\rm crit}$ and $B_{\rm crit}$ are free parameters determined
from the data, as discussed in \S5.

We further parametrize the wind velocity $\va$ along field lines at
the Alfv\'en radius $\ra$ as:
\begin{equation} 
  \label{eq:windvel}
  \va \equiv \vp\left(\frac{\ra}{R}\right)^q
\end{equation}
where $\vp$ is some constant with dimensions of velocity defined on
the stellar surface, and $q$ is the parametric exponent.  $\va$ is
empirically unknown, and must be specified.  We follow K88 in assuming
that $\va$ is proportional to the escape velocity at the Alfv\'{e}n
surface: $\va = K_V \sqrt{GM/\ra}$, where $K_V$ is some dimensionless
constant scaling factor.  Thus we have $\vp = K_V \sqrt{GM/R}$ and $q
= -1/2$.  While other choices of velocity may also be argued for
(e.g., see M84), here we stick to K88's choice of the escape velocity
(comparable to the thermal velocity adopted by M84 for `slow
rotators') to hew as closely as possible to the models currently used.

Finally, we assume that the mass loss rate $\mdot$, saturation field
strength $B_{\rm crit}$, critical angular velocity for saturation
$\Omega_{\rm crit}$, and the velocity scaling factor $K_V$ are all
constant with time and the same for stars of all masses.  While this
may well be an over-simplification, we have very little observational
or theoretical guidance on how to fix the time- and mass-dependence of
these quantities.  As such, we {\it a priori} ignore such potential
dependencies in our quest for the simplest physically motivated theory
to compare to observations.  Note, in particular, that this means we
do {\it not} assume a mass-dependent scaling of $\Omega_{\rm crit}$
from the outset, in contrast to most current formulations of angular
momentum evolution \citep[e.g.,][]{1996ApJ...462..746B,
  1997ApJ...480..303K, 2000ApJ...534..335S}.  Any discrepancies that
arise in the comparison of our model to the data will then {\it
  motivate} an examination of such dependencies {\it a posteriori}, as
discussed in \S\S7 and 8.

Now, assuming a spherical Alfv\'enic surface and field-enforced
corotation of the gas out to $\ra$ at the stellar angular velocity,
the rate at which the star loses angular momentum is given by:
\begin{equation}
  \label{eq:dJdt}
  \frac{dJ}{dt} = -\frac{2}{3} \mdot \Omega \ra^2
\end{equation} 
To make progress, we note that by the definition of the Alfv\'en
radius, the wind velocity $\va$ at $\ra$ must equal the Alfv\'en
velocity there: $\va \equiv B_A/\sqrt{4\pi\rho_A}$, where $B_A$ and
$\rho_A$ are the field strength and wind density at $\ra$.  Moreover,
$B_A = \bs(R/\ra)^2$ for a radial field; mass continuity along field
lines demands $\rho_0 v_0/\bs = \rho_A \va/B_A$; and the mass loss
rate is given by $\mdot = 4\pi\ra^2\rho_A\va$.  Inserting these into
equations [2--4], with the parametric exponents $a=1.5$ and $q=-1/2$,
finally yields the rate of angular momentum loss with radial fields to
be:
\begin{eqnarray}
  \label{eq:final}
  \frac{dJ}{dt} = -\,\mathcal{C} \left[\Omega \left(\frac{R^{16}}{M^2}\right)^{1/3}\right] \,\,\,\,\,\,\,\,\,\, {\rm for} \,\,\,\,\, \Omega \ge \Omega_{\rm crit} {\nonumber}\\
  {\nonumber}\\
  \frac{dJ}{dt} = -\,\mathcal{C} \left[\left(\frac{\Omega}{\Omega_{\rm crit}}\right)^{4}\Omega\left(\frac{R^{16}}{M^2}\right)^{1/3}\right] \,\,\,\,\,\,\,\,\,\, {\rm for} \,\,\,\,\, \Omega < \Omega_{\rm crit} {\nonumber}\\
  {\nonumber}\\
  {\rm with} \,\,\,\,\, \mathcal{C} \equiv \frac{2}{3}\left(\frac{B_{\rm crit}^{\,8}}{G^{\,2} K_V^{\,4} \mdot}\right)^{1/3} \,\,\,\,\,\,\,\,\,\,\,\,\,\,\,\,\,\,\,\,\,\,\,\,\,\,\,\,\,\,\,\,\,\,\,\,\,\,\,\,\,\,\,\,\,\,\,
\end{eqnarray}         
The terms within square brackets in the expressions for $dJ/dt$ affect
the time-evolution and mass-dependence of the angular momentum loss
rate, while the constant $\mathcal{C}$, comprising terms that are
(assumed to be) star- and time-independent, affects the global scaling
of $dJ/dt$.

There are two noteworthy points about equation [5].  First, it
includes a strong dependence on the stellar radius $R$.  If we had
instead followed K88 in parametrizing the magnetic {\it flux} in terms
of the angular velocity, the dependence on the stellar radius would
have decreased by a factor of $R^{16/3}$, i.e., $dJ/dt$ would have
become entirely independent of the radius (as shown explicitly by
K88's equation [10], using his exponent $n$=2 for radial fields).
This insensitivity to $R$, contrary to our equation, accounts for why
previous studies based on the K88 formulation have not identified the
evolution of the stellar radius as being vital to understanding
magnetic braking\footnote{K88 discusses the effects of varying field
  geometries; while we only discuss radial fields here, the
  fundamental point is that for $any$ specified field geometry, K88's
  formulation in terms of magnetic flux yields a much weaker
  dependence of $dJ/dt$ on stellar radius than our formulation in
  terms of magnetic field strength.}.

Second, the rate of angular momentum loss is also very sensitive to
the saturation threshold $\Omega_{\rm crit}$, with the loss rate
declining rapidly once the stellar angular velocity decreases below
this limit. This combined dependence on stellar radius and
$\Omega_{\rm crit}$ implies the following.

\section{Trends in the Model Spin-up and Spin-down} 

\begin{figure}
    \includegraphics[width=16.6cm,clip=]{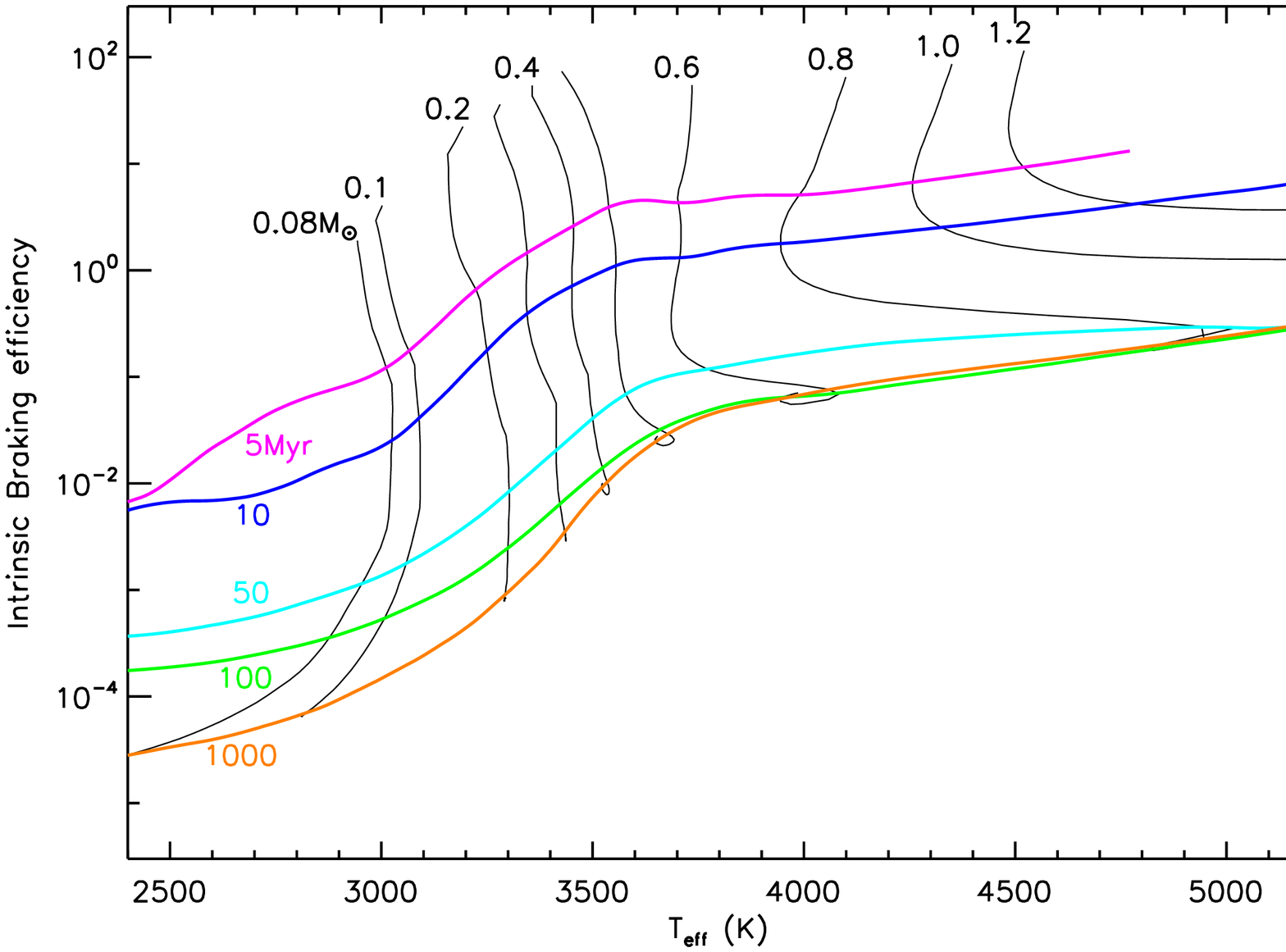}
    \caption{\label{fig:efficiency}Relative intrinsic braking
      efficiency $R^{16/3}M^{-2/3}$, black lines show evolutionary
      tracks for [.08, 0.1, 0.2, 0.3, 0.4, 0.5, 0.6, 0.8, 1,
      1.2]\,M$_{\odot}$, colored lines show isochrones at ages [5, 10,
      50, 100, 1000] Myr.}
\end{figure}

Consider first the effect of stellar radius alone.  During PMS
evolution, contraction drives spin-up, which halts when the star
reaches a stable radius on the MS.  Concurrently, spin-down due to
angular momentum loss decreases with smaller stellar radius, by
equation [5], during both PMS and MS phases.  Since a less massive
star has a smaller radius at a given age and also arrives later on the
MS, it follows that rotation will tend to be faster, and the spin-down
timescale longer, with decreasing mass at any specified age.
Fig.\,\ref{fig:efficiency} illustrates the radius-dependence of the
loss (braking) rate $dJ/dt$, given by the term $(R^{16}/M^2)^{1/3}$ in
equation [5], for low-mass stars over a range of ages \citep[we use
mass-radius-age relationships from the theoretical evolutionary tracks
by ][hereafter BCAH98]{1998AA...337..403B}.  We call this term the
``intrinsic braking efficiency'' (since it depends only on the
evolution of the stellar structure, which we assume here is
rotation-independent).  We see that {\it (1)} overall, the intrinsic
braking efficiency falls off with decreasing mass at any given age,
and {\it (2)} the falloff with diminishing mass steepens with age, as
solar-type stars arrive on the MS and cease contraction upon the
formation of a radiative core (by $\sim$30 Myr for 1\,M$_{\odot}$)
while fully convective VLMS continue contracting to much smaller radii
and arrive on the MS at increasingly later ages (from $\sim$200 Myr at
0.3\,M$_{\odot}$ to $\sim$600 Myr at 0.1\,M$_{\odot}$).  {\it This
  behaviour of the intrinsic braking efficiency is a fundamental
  ingredient in the evolution of angular momentum in low-mass stars,
  and crucial for understanding the observed mass-dependence of their
  rotation periods and the very fast rotation of VLMS at a few Gyr}.

We now further assume {\it solid body rotation} ($J = 2MR^2\Omega/5$)
in all cases.  This is a good approximation for fully convective
objects, but may break down temporarily when a radiative core develops
(due to `core-envelope decoupling'); we discuss this issue further in
\S8.  For now, this assumption implies that once a star arrives on the
MS, it spins down as (integrating equation [\ref{eq:final}] for any
given stellar mass at its fixed MS radius):
\begin{eqnarray}
  \label{eq:integration}
  \frac{\Omega(t)}{\Omega_0} = {\rm e}^{-(t-t_0)/t_{S}} \,\,\,\,\, , \,\,\,\,\, t_{S} \equiv\left[\mathcal{C}\frac{5}{2MR^2}\left(\frac{R^{16}}{M^2}\right)^{1/3}\right]^{-1}\,\,\,\,\, (t_0 \leq t < t_{\rm crit}\,\, {\rm : saturated}){\nonumber}\\
  {\nonumber}\\
  \frac{\Omega(t)}{\Omega_{\rm crit}} = \left[\frac{(t-t_{\rm crit})}{t_{U}} + 1\right]^{-1/4} \,\,\,\,\, ,\,\,\,\,\, t_{U} \equiv \left[4\mathcal{C}\frac{5}{2MR^2}\left(\frac{R^{16}}{M^2}\right)^{1/3}\right]^{-1}\,\,\,\,\, (t_{\rm crit}\leq t \,\,\,\,\, {\rm : unsaturated})
\end{eqnarray}

Here $t_{S}$ and $t_{U}$ are the MS spin-down timescales in the {\bf
  s}aturated and {\bf u}nsaturated domains respectively; $t_0$ is the
age at which the star arrives on the MS (set by stellar evolution);
$\Omega_0$ is its angular velocity at that time (set by a combination
of PMS spin-up and spin-down, with the former dominating since the
average timescale for contraction during PMS evolution is shorter than
that for spin-down; see \S4.1 below); and $t_{\rm crit}$ is the
subsequent age at which the star slows to below the critical rate
$\Omega_{\rm crit}$ (with $t_{\rm crit}$ determined by setting
$\Omega(t) = \Omega_{\rm crit}$ in the saturated equation $\Rightarrow
t_{\rm crit} = t_0 +t_{S}\,{\rm ln}\left[\Omega_0/\Omega_{\rm
    crit}\right]$).  Thus a star arrives on the MS spinning rapidly,
in the saturated regime, and then slows down exponentially quickly to
$\Omega_{\rm crit}$; thereafter the spin-down rate diminishes to a
very weak power-law, and the star remains within a factor of a few of
$\Omega_{\rm crit}$ for the rest of its MS lifetime.  For a star of a
given mass (and hence MS radius), the constant $\mathcal{C}$
determines, via $t_{S}$ and $t_{U}$, precisely how quickly the
unsaturated regime is achieved and how close the star remains to
$\Omega_{\rm crit}$ thereafter.

\subsection{Comparison to Skumanich Law}

In our model, stars on the MS first spin down exponentially rapidly to
$\Omega_{\rm crit}$; thereafter the spin-down rate diminishes
significantly, with the decrease in angular velocity with time
asymptotically approaching a weak power-law: $\Omega(t) \propto
t^{-1/4}$.  On the other hand, observed rotation rates in solar-type
stars from the age of the Pleiades ($\sim$100 Myr) to the Sun seem to
approximately follow the empirical Skumanich law
\citep{1972ApJ...171..565S}: $\Omega(t) \propto t^{-1/2}$.  Is our
model compatible with the latter?

\begin{figure}
    \includegraphics[width=16.6cm,clip=]{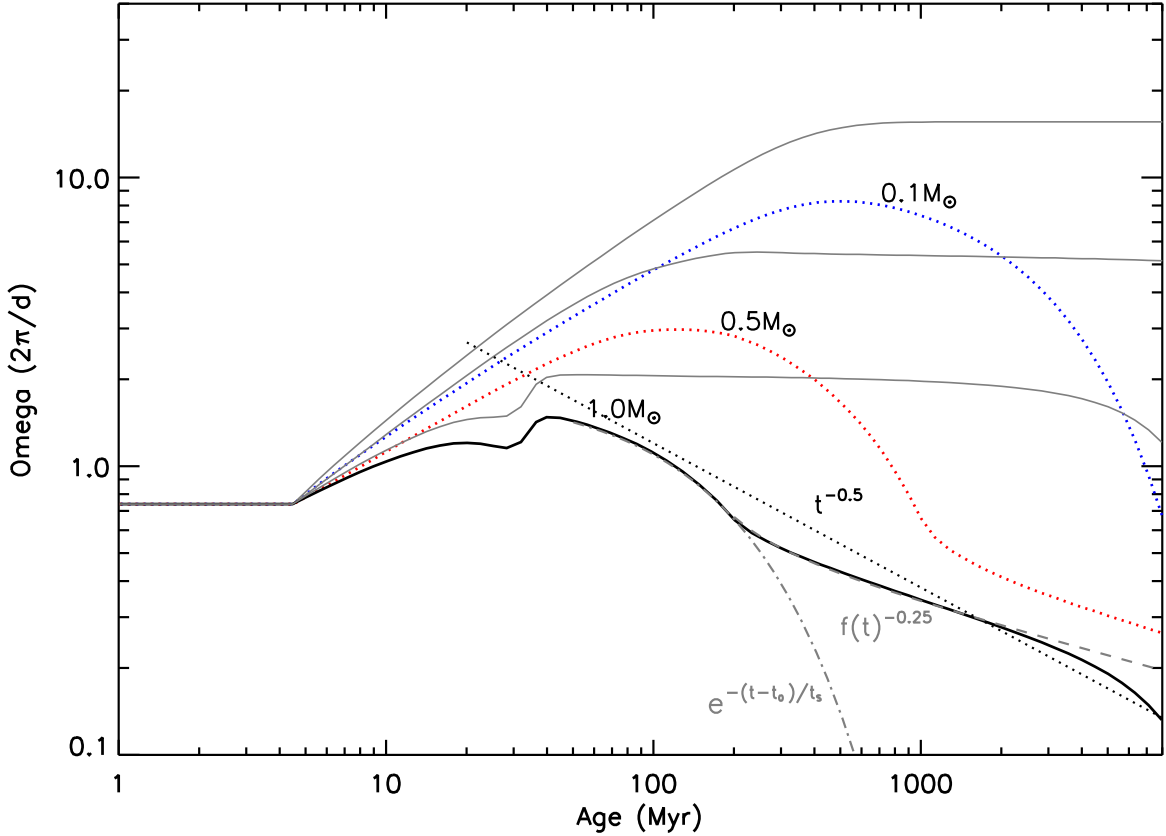}
    \caption{\label{fig:age}Angular velocity evolution according to
      Eq.\,\ref{eq:integration} for three different model stars; black
      solid line: 1\,M$_{\odot}$; red dotted line: 0.5\,M$_{\odot}$;
      blue dotted line: 0.1\,M$_{\odot}$. The Skumanich braking law
      $t^{-0.5}$ is shown for comparison (black dotted line). For the
      1\,M$_{\odot}$ case, we overplot the braking laws for saturated
      regimes; saturated case, $t < t_{\rm crit}$: $e^{-(t-t_0)/t_S}$
      (grey dash-dotted line), and unsaturated case, $t > t_{\rm
        crit}$: $[(t-t_{\rm crit})/t_U]^{-1/4}$ (grey dotted
      line). Grey solid lines show hypothetical angular velocity
      evolution in the absence of any braking for the three model
      masses considered.}
\end{figure}

We examine this in Fig.\,\ref{fig:age}, where we plot our model
predictions for angular velocity as a function of time for stars of
mass 1, 0.5 and 0.1 M$_{\odot}$.  To best illustrate the differences
between the masses, we have adopted the same initial rotation period
for all three: $\sim$8d (within the range of initial periods observed
in very young clusters; see \S7).  The velocities are scaled such that
the 1 M$_{\odot}$ curve replicates the rotation period of the
present-day Sun (specifically, we have used the best-fit values of
$\Omega_{\rm crit}$ and $\mathcal{C}$ for our model, the choice of
which is described in \S\S5 and 8.1.1).  For all three stars, we
overplot the rotation curves expected for pure spin-up during the PMS
phase.  For the 1 M$_{\odot}$ case, we also overplot the individual
$\Omega(t)$ predicted by our model for the saturated and unsaturated
regimes on the MS (equation [6]), as well as the Skumanich law.

Four facts are immediately apparent.  First, for almost their entire
PMS lifetimes ($\sim$30, 150 and 600 Myr for 1, 0.5 and 0.1
M$_{\odot}$ respectively), spin up dominates in these stars.  This
illustrates our earlier statement that the angular velocity at which
low-mass stars arrive on the MS is mainly set by spin-up due to PMS
contraction, since the contraction timescales are shorter in the PMS
phase than the spin-down timescales.  Nevertheless, we see that there
is some contribution from angular momentum loss during this phase as
well: stars arrive on the MS spinning somewhat slower (by a factor
$\lesssim$2) than predicted by PMS spin-up alone.  The curves also
illustrate our earlier point that, for a given initial rotation
period, higher mass stars always rotate slower than lower mass ones
and spin down faster, due to a combination of larger radius, earlier
arrival on the MS and higher intrinsic braking efficiency.

Second, for ages from $\sim$30 Myr to 2 Gyr, the Skumanich curve is in
close agreement with the {\it average} trend in angular velocity for
our 1 M$_{\odot}$ model, where the latter is a combination of
exponential decay in the saturated regime (up to $\sim$200 Myr) and
power-law decay in the unsaturated domain (at $>$200 Myr).
Specifically, our predicted angular velocities at Pleiades and Hyades
ages (100 and 650 Myr respectively) deviate by only 20--30\% from the
Skumanich curve\footnote{This relatively small difference is even less
  significant considering that Skumanich (1972) used the {\it mean}
  rotation rates in these two clusters to construct his fit, while the
  1 M$_{\odot}$ model we plot here corresponds to only one choice of
  initial rotation period out of the range observed in young clusters.
  For solar-mass stars, the effect of the initial period continues to
  be significant at 100-200 Myr (i.e., Pleiades ages), though it is
  negligible by the age of the Hyades (see Fig. 4, and discussion in
  \S8).}, and by much less from 650 Myr to 2 Gyr.  In other words, the
Skumanich law is a very good {\it linear} approximation (in a log-log
plot) to our model from 100 Myr to 2 Gyr.

Third, the Skumanich curve is a very good match to our 1 M$_{\odot}$
model from 2--8 Gyr, while our $\Omega(t) \propto t^{-1/4}$ curve lies
above both after $\sim$3 Gyr. The reason is that solar-mass stars have
already begun evolving {\it off} the MS by 3 Gyr, becoming larger at
later ages (the current solar radius is $\sim$10\% greater than its MS
value).  While our full model incorporates this radius change, our
power-law curve for the unsaturated regime is valid {\it only on the
  MS} (as stated in the derivation of equation [6]).  The increasing
radius at $>$3 Gyr spins down solar-mass stars faster than the
$\Omega(t) \propto t^{-1/4}$ expected for a constant MS radius, and
makes our 1 M$_{\odot}$ model nearly identical to the $\Omega(t)
\propto t^{-1/2}$ empirical Skumanich curve at these ages.

Thus, the Skumanich curve for solar-types is better understood as a
{\it mean} fit to rotation data from the Pleiades to the Sun, born of
three distinct physical phenomena: exponential spin-down in the
saturated regime on the early MS (Pleiades ages); power-law spin-down
in the unsaturated regime on the mid- to late MS ($\gtrsim$ Hyades
ages); and expansion in the early post-MS (present-day Sun).  Our
model explicitly accounts for each of these processes, and in doing so
yields good overall agreement with the empirical Skumanich law for
solar masses.

Finally, Fig.\,\ref{fig:age} also shows that a Skumanich power law is
{\it not} a good match to our model predictions for VLMS.  For a fixed
initial rotation period, these stars arrive on the MS spinning
considerably faster than their solar-mass counterparts (because of
their longer PMS lifetimes and much smaller MS radii, as discussed
earlier); consequently, they remain in the saturated, exponential
spin-down regime for much longer on the MS, and a $t^{-1/2}$ power law
is far too shallow to fit their MS angular velocity evolution up to
ages of a few to several Gyr.  As we show shortly, our model
accurately reflects the observed behaviour of VLMS.

\section{Choice of $\Omega_{\rm crit}$ and $\mathcal{C}$} 

The specific exponent for the radius-dependence in equation [5] is
fixed by our choices of field geometry (radial) and Alfv\'{e}n
velocity (proportional to escape velocity), which are justified on the
grounds of being the simplest possibilities (though we reexamine their
validity later).  A similar a priori choice of $\Omega_{\rm crit}$ and
$\mathcal{C}$, however, is much harder.  From a theoretical
perspective, $\Omega_{\rm crit}$ may ultimately depend on the
convective turnover timescale ($\tau_{\rm conv}$)
\citep{1997ApJ...480..303K}.  However, $\tau_{\rm conv}$ is
ill-defined, poorly constrained by theory and data, and possibly
strongly time- and mass-dependent \citep{1996ApJ...457..340K}, so
invoking it a priori only introduces more free parameters (we do
examine its importance a posteriori, and find it may indeed play a
role).  Observationally, VLMS in the field (M dwarfs) are mostly
saturated, but where $\Omega_{\rm crit}$ occurs and saturation ends is
unclear \citep[because the $v\,\sin{i}$ fall below detectable
limits;][]{2007A&A...467..259R}.  Conversely, field solar-mass stars
are predominantly {\it un}saturated with very slow rotation, again
making $\Omega_{\rm crit}$ hard to determine.

Similarly, the factor $\mathcal{C}$ involves the poorly known
quantities $B_{\rm crit}$, $K_V$ and $\mdot$. Observations of
saturated field M dwarfs as well as very young T Tauri stars indicate
$B_{\rm crit}$ of a few kG; nevertheless, the field star data in this
regard are very limited and poorly constrained
\citep{1996IAUS..176..237S}.  $K_V$ must a priori be of order unity,
if our choice of escape velocity is to be reasonably valid, but from
an observational perspective it is unknown.  Finally, the mass-loss
rate $\mdot$ is usually assigned the present-day solar value of
$\sim$10$^{-14}$ $M_{\odot}$yr$^{-1}$, but in reality is also very
poorly determined (or not at all) for other stellar masses and ages.
Recent simulations, for instance, suggest $\mdot$ may be orders of
magnitude larger in VLMS \citep{2011MNRAS.412..351V}.  In this sense,
our assumption of a fixed $\mdot$ for all stars and ages (as usually
assumed in studies of rotational evolution) represents a {\it globally
  and temporally averaged} mass loss-rate, but we do not know what the
actual value of this mean is.

Given these theoretical and observational uncertainties, we consider
$\Omega_{\rm crit}$ and $\mathcal{C}$ to be the two free parameters in
our model, which are determined as follows.  Given an initial
distribution of angular momenta, a specified pair of values
[$\Omega_{\rm crit}$, $\mathcal{C}$] (mass and time-independent in our
simple theory) uniquely fixes the shape of the period-mass curve and
the absolute scaling of the periods at {\it every} subsequent age.  We
therefore evolve an {\it observed} sample of rotation periods at a
very young age (which serves as our estimate of the initial angular
momentum distribution; \S7) forward in time to the age of the Sun
(i.e., a few Gyrs), using a range of values for $\Omega_{\rm crit}$
and $\mathcal{C}$.  The [$\Omega_{\rm crit}$, $\mathcal{C}$] pair that
best fits {\it both} the present-day solar rotation period {\it and}
the rotation-mass distribution of field stars at roughly the same age
then represents our best estimate of these two parameters (\S8.1.1).
Note that the individual empirical uncertainties in $B_{\rm crit}$,
$K_V$ and $\mdot$ are unimportant, since they are subsumed within the
single quantity $\mathcal{C}$, making this a simple problem of fixing
two unknowns with two observations.

We then {\it test} our theory by using this best-fit [$\Omega_{\rm
  crit}$, $\mathcal{C}$] pair to similarly generate the mass-period
distribution at various {\it other} ages, and comparing to
observations (\S8.1.2).  We also test the theory by comparing its
predictions to the observed lifetimes of magnetically-driven activity
in low-mass stars (\S8.3).  Finally, we examine the plausibility of
our inferred best-fit $\Omega_{\rm crit}$ (\S8.2) and $\mathcal{C}$
(\S9) as a separate constraint on the model.  These tests in turn
provide some deeper physical insights into the processes involved in
rotation regulation.

Below, we first discuss the data we use, before going on to
comparisons with our model.

\section{Rotation and Activity Data, and Stellar Mass Determination}

In our analysis of the data (\S\S7 and 8), masses for the observed
stars have been inferred as follows.  In the vast majority of cases,
we have used the mean distance to the star-forming region or open
cluster, or the known distance to the individual stars from parallax
measurements, together with extinction data, to convert the apparent
magnitude in a selected photometric band ($I_C$, $J$ or $K$) to an
absolute magnitude.  Masses are then derived from mass-magnitude
relationships: either theoretical ones from the BCAH98 evolutionary
tracks for the adopted age of the region/cluster, or empirical ones
for low-mass stars on the MS \citep{2000A&A...364..217D,
  2008ApSS.314...51X}.  In the handful of cases where this is onerous,
T$_{\rm eff}$ are calculated from either spectral type--T$_{\rm eff}$
or color--T$_{\rm eff}$ empirical relationships, and masses from the
mass-T$_{\rm eff}$ relationship supplied by the BCAH98 tracks for the
age of the cluster.

The BCAH98 tracks we employ are actually a concatenation of two sets
of models: those using a convective mixing-length parameter of
${\alpha}_{\rm mix} = 1.0$, appropriate for masses $\lesssim$0.6
M$_{\odot}$, and those with ${\alpha}_{\rm mix} = 1.9$ (the value
required to fit the Sun), appropriate for masses $>$0.6 M$_{\odot}$
(see discussion in Baraffe et al. 2002).  The concatenated set yields
a smooth mass-magnitude relationship spanning the two $\alpha_{\rm
  mix}$ regimes for any specified age.  We further note that the
BCAH98 tracks, which do not incorporate the formation of photospheric
dust, are appropriate for the 0.1--1 M$_{\odot}$ range investigated in
this paper (chemical equilibrium calculations indicate that dust
formation becomes important at T$_{\rm eff}$ $\lesssim$ 2500K
\citep{2001ApJ...556..357A}, significantly lower than the temperatures
of $\gtrsim$ 2800K expected for stellar masses of 0.1--1 M$_{\odot}$
over the $\sim$1 Myr -- 10 Gyr age range considered here).

Specifically, for the individual populations we have examined,
rotation data and stellar masses are obtained as follows.

\noindent {\it Orion Nebula Cluster (ONC)}: Stellar periods are taken
from \citet{2002A&A...396..513H} (who have compiled data from both
their own study and from Herbst et al. 2000 and Stassun et al. 1999).
Photometry and extinctions for these stars are from Hillenbrand (1997;
used by Herbst et al. 2002 as well).  Absolute $I_C$-band magnitudes
($M_{I_C}$) are computed from the observed $I_C$ and $A_V$ (latter
converted to $A_{I_C}$ assuming a normal extinction law: $R_V \equiv
A_V / E(B-V) = 3.1 \Rightarrow A_{I_C} = 0.60 A_V$
\citep{1998ApJ...500..525S}), and adopting a mean distance to the ONC
of $d$ = 450 pc (Herbst et al. 2002).  Masses are then derived from
the BCAH98 mass--$M_{I_C}$ relationships, assuming a mean age of 1 Myr
for the ONC \citep{2009IAUS..258..363I}.

\noindent {\it NGC 2264}: Rotation and photometric data are from
\citet{2005A&A...430.1005L}.  Masses are calculated as above, from the
BCAH98 mass-$M_{I_C}$ relationships, using the mean $E(B-V)$ = 0.55
mag and $d$ = 760 pc adopted by \citet{2005A&A...430.1005L} and
assuming a mean age of 2 Myr \citep{2009IAUS..258..363I}.  Note that
Lamm et al.\,\,use \citet{1997MmSAI..68..807D} evolutionary tracks to
infer a mean age of 0.5 Myr for the ONC and 1 Myr for NGC 2264, half
the values we adopt for these two regions.  However, as Lamm et
al.\,\,note, the latter tracks give systematically smaller ages
compared to others; since we use BCAH98 models instead, we adopt the
larger ages provided by \citet{2009IAUS..258..363I} based on
comparisons to the same models.

\noindent {\it M50}: Rotation and photometric data are from
\citet{2009MNRAS.392.1456I}.  Masses are again from the BCAH98
mass-M$_{I_C}$ relationships, using the mean $A_{I_C}$ = 0.25 mag, $d$
= 1000 pc, and age = 150 Myr adopted by Irwin et al. (who also employ
the same method to derive masses).

\noindent {\it Praesepe / Hyades}: Since the Praesepe and Hyades
clusters are nearly coeval with an age of $\sim$600--650 Myr
\citep{2011MNRAS.413.2218D, 2009IAUS..258..363I}, we follow
\citet{2009IAUS..258..363I} in lumping them into a single population
with an adopted mean age of 650 Myr.  Rotation and photometric data
for Praesepe are from \citet{2007MNRAS.381.1638S},
\citet{2011MNRAS.tmp..261S} and \citet{2011MNRAS.413.2218D}, and for
Hyades from \citet{1987ApJ...321..459R}, \citet{1995PASP..107..211P}
and \citet{2011MNRAS.413.2218D}.  The Delorme et al. survey makes up
the bulk of the data.

Crucial to our mass determination here is the fact that by $\sim$600
Myr, all stars in the mass range 0.1--1 M$_{\odot}$ have arrived on
the MS.  \citet{1987ApJ...321..459R} and \citet{1995PASP..107..211P}
focus on a relatively small number of solar-type stars in the Hyades,
and supply ($B-V$) colors for their samples; we convert the latter to
T$_{\rm eff}$ using the empirical MS color--T$_{\rm eff}$ relationship
compiled by \citet{1995ApJS..101..117K}, and thereby derive masses
from the BCAH98 mass-T$_{\rm eff}$ relationship for 650 Myr.  For
Praesepe, \citet{2007MNRAS.381.1638S} and \citet{2011MNRAS.tmp..261S}
have determined masses by calculating $M_{I_C}$ and $M_J$ from their
$I_C$ and $J$ photometry respectively (for a mean distance of
170\,pc), and then applying the BCAH98 mass-magnitude relationships
for an age of 630 Myr (essentially identical to our adopted 650 Myr).
For MS M dwarfs (stars $\lesssim$0.6 M$_{\odot}$, comprising nearly
the entire sample in the latter two surveys),
\citet{2000A&A...364..217D} show that the BCAH98 models are a very
good match to the tight {\it empirical} MS mass-magnitude
relationships in the near-infrared (and argue that the same is likely
true in the $I_C$-band too).  Consequently, we adopt these masses
unchanged.

Finally, for the \citet{2011MNRAS.413.2218D} sample, we calculate
$M_K$ from their $K$-band photometry, assuming a distance of 170\,pc
to Praesepe and 45\,pc to Hyades\footnote{\citet{2011MNRAS.413.2218D}
  provide parallaxes for some of their Hyades sample, but not for the
  majority; we therefore use the mean distance to the Hyades in all
  cases for uniformity.  The parallactic distances they do provide are
  consistent with a small scatter around our mean $d$ = 45 pc.}, and
then derive masses from the empirical MS mass-$M_K$ relationships of
\citet{2000A&A...364..217D} appropriate for $\lesssim$0.6 M$_{\odot}$)
and \citet{2008ApSS.314...51X} applicable to stars $\sim$0.6--1
M$_{\odot}$: for 0.1--0.6 M$_{\odot}$. The Xia et al. mass-$M_K$ fit
is nearly indistinguishable from that of Delfosse et
al.)\footnote{Delorme et al. supply $K_{2MASS}$, while the two
  empirical mass-magnitude relationships we employ use $K_{CIT}$.  The
  difference between the two filters is negligible for our purposes,
  however, over the 0.1--1 M$_{\odot}$ range of interest here, and we
  ignore it.}.

It is worth noting that the masses \citet{2011MNRAS.413.2218D} find
for their sample are in some cases significantly at odds with ours,
with discrepancies of up to 30\% at the lowest masses.  This is
because they derive mass from ($V-K$) color instead of from $M_K$.
The $V$-band is known to be severely affected by metallicity
variations, unlike the $JHK$-bands (BCAH98; Delfosse et al. 2000; Xia
et al. 2008); thus, on the MS, the mass--($V-K$) relationship evinces
much greater scatter than the very tight mass-$M_K$ relationship
\citep{2000A&A...364..217D}, and the latter yields far better mass
estimates.  This has important consequences for understanding
rotational evolution as a function of mass, as we point out in
\S8.1.2.

\noindent {\it Field M dwarfs}: Rotation periods and photometric data,
for both young disk and old disk/halo field M dwarfs, are from
\citet{2011ApJ...727...56I}.  These authors use $M_K$ (calculated from
literature $K$-band photometry combined with parallactic distances) to
derive masses from the empirical MS M dwarf mass-$M_K$ relationship of
\citet{2000A&A...364..217D}.  Since this is our preferred method for
field M dwarfs (as discussed above), we adopt these masses unchanged.
We note that the old disk/halo stars may have lower metallicities than
the young disk population; however, this is unlikely to skew the mass
estimates, given the insensitivity of the \citet{2000A&A...364..217D}
relationship in the $K$-band to metallicity (see above).

Lastly, we also compare our model predictions to empirical activity
lifetimes for MS M dwarfs.  We obtain the mean activity lifetime for
each M spectral sub-class from the large survey by
\citet{2008AJ....135..785W}.  The $T_{\rm eff}$ corresponding to each
sub-type is determined from the MS spectral type--T$_{\rm eff}$
calibrations of \citet{1995ApJS..101..117K} and
\citet{2004AJ....127.3516G}; corresponding masses are then derived
from the BCAH98 MS mass-T$_{\rm eff}$ relationship.

\section{Choice of Initial Conditions and Disk-Locking}

In order to compare our theory of angular momentum evolution to
stellar data at various epochs, we must first specify the initial
distribution of angular momenta in our model.  Observed rotation rates
in very young star-forming regions (SFRs) provide our best estimate of
this initial condition.  The ONC (at $\sim$1 Myr) and NGC 2264 (at
$\sim$2 Myr), with the most extensive data on rotation in newborn
low-mass stars, are currently the SFRs of choice in this regard.  To
maximise the sample size, we combine the data for the two regions as
follows.

Most current models of angular momentum evolution assume a period of
`disk-locking' during the initial disk accretion phase, wherein the
angular velocity of the star is held constant by star-disk
interactions \citep{1994ApJ...429..781S, 2008ApJ...687.1323M}.  While
there is considerable debate about the mechanism, efficiency, lifetime
and mass-dependence of this phenomenon, it seems to play a role in at
least some significant fraction of low-mass stars (see review by
\citet{2007prpl.conf..297H} and extensive references therein).  In
particular, many young accreting solar-type stars are observed to
rotate much slower than possible in the presence of only spin-up due
to gravitational contraction, indicating some source of braking;
modeling the evolution of these slow rotators from the PMS to the
zero-age MS also seems to require substantial braking during the early
PMS phase \citep{2007prpl.conf..297H}.  Disk-locking provides such a
mechanism.  Furthermore, the locking timescales implied for solar-mass
stars by such modeling is $\sim$5--10 Myr, consistent with the
observed accretion timescale in these stars.  For VLMS, disk-locking
has been less scrutinized, but there is some evidence that it operates
in these stars as well -- accreting VLMS (and brown dwarfs) seem to
rotate preferentially slower than non-accreting ones
\citep{2004AA...419..249S, 2005MmSAI..76..303M} -- albeit perhaps less
efficiently than in solar-mass stars \citep{2005A&A...430.1005L}.
Moreover, the accretion timescale in VLMS is also 5--10 Myr, similar
to that in solar-types \citep{2005ApJ...626..498M}.

Under the circumstances, we assume the simplest scenario: disk-locking
for the first 5 Myr, for all stars within the 0.1--1 M$_{\odot}$ range
of interest here.  The same condition is adopted by
\citet{2009IAUS..258..363I}.  To impose this constraint on our initial
conditions -- given by the observed rotation rates in the ONC and NGC
2264 -- we simply assign the stars in these SFRs the {\it radii}
predicted by the BCAH98 tracks at 5 Myr for their derived masses,
while keeping their {\it rotation periods} fixed at the observed
values.  This mimics gravitational contraction from 1 to 5 Myr (for
the ONC), or from 2 to 5 Myr (for NGC 2264), at constant angular
velocity, which is what disk-locking till 5 Myr means.  The two
samples are then merged, with the combined dataset representing the
period and radius distribution expected at the end of the disk-locking
phase for our initial conditions; this forms the starting point for
our model evolution beyond 5 Myr.  The period distribution in this
dataset is plotted in Fig.\,\ref{fig:periods} (left panel); most of
the stars lie between 0.8 and 10\,d, with a few as slow as 20 to
30\,d.  A caveat, mentioned earlier, is that the data may be biased by
selection effects: we may be missing very rapid and/or very slow
rotators.  This must be clarified by future surveys.

\section{Results} 

\subsection{Radius Dependent Evolution, and Transient Core-Envelope
  Decoupling}

\subsubsection{Best-Fit $\Omega_{crit}$ and $\mathcal{C}$} 

\begin{figure}
\mbox{
  \parbox{.33\textwidth}{
    \includegraphics[bb = 125 10 625 430, width=.33\textwidth]{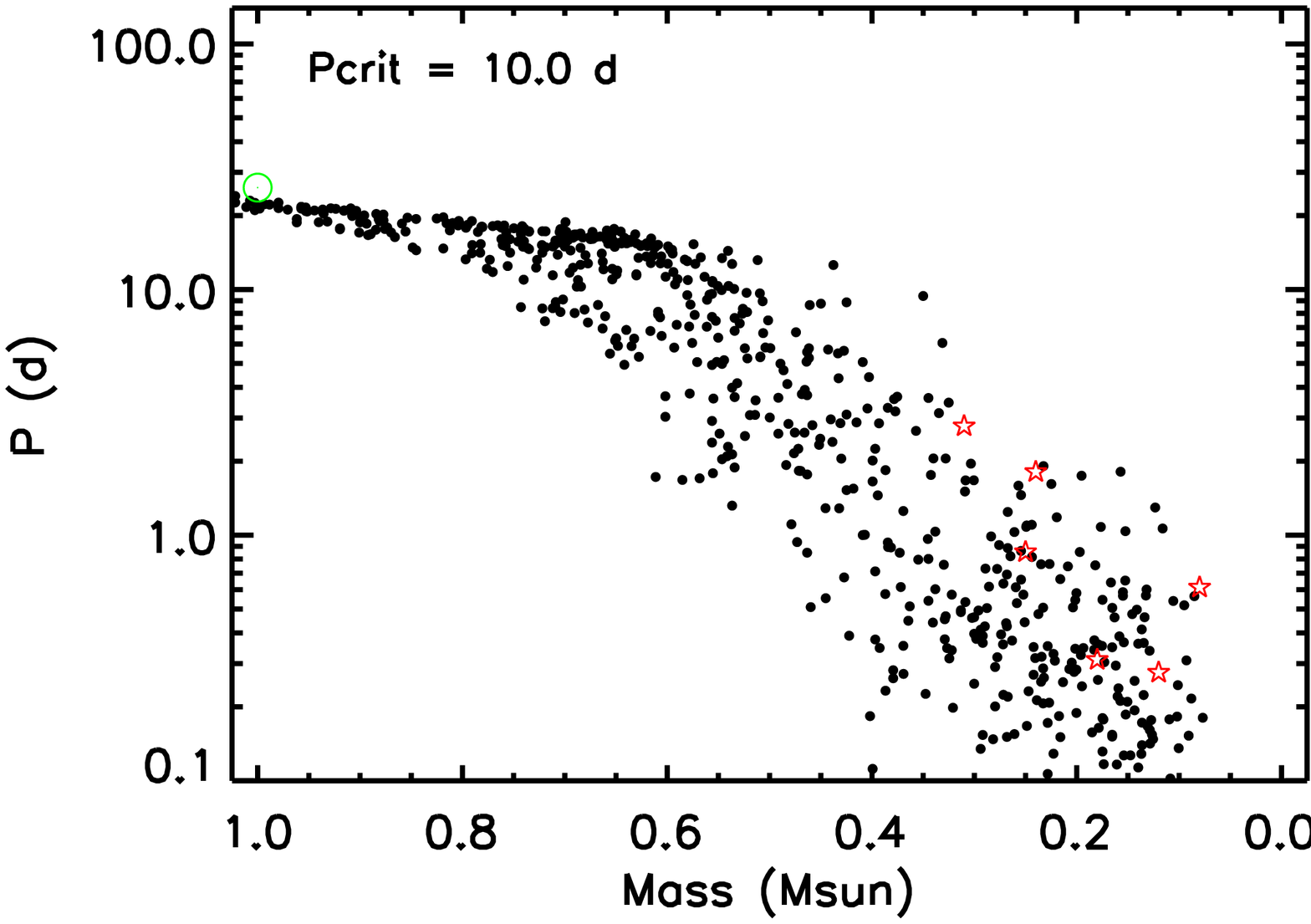}}
  \parbox{.33\textwidth}{
    \includegraphics[bb = 125 10 625 430, width=.33\textwidth,clip=]{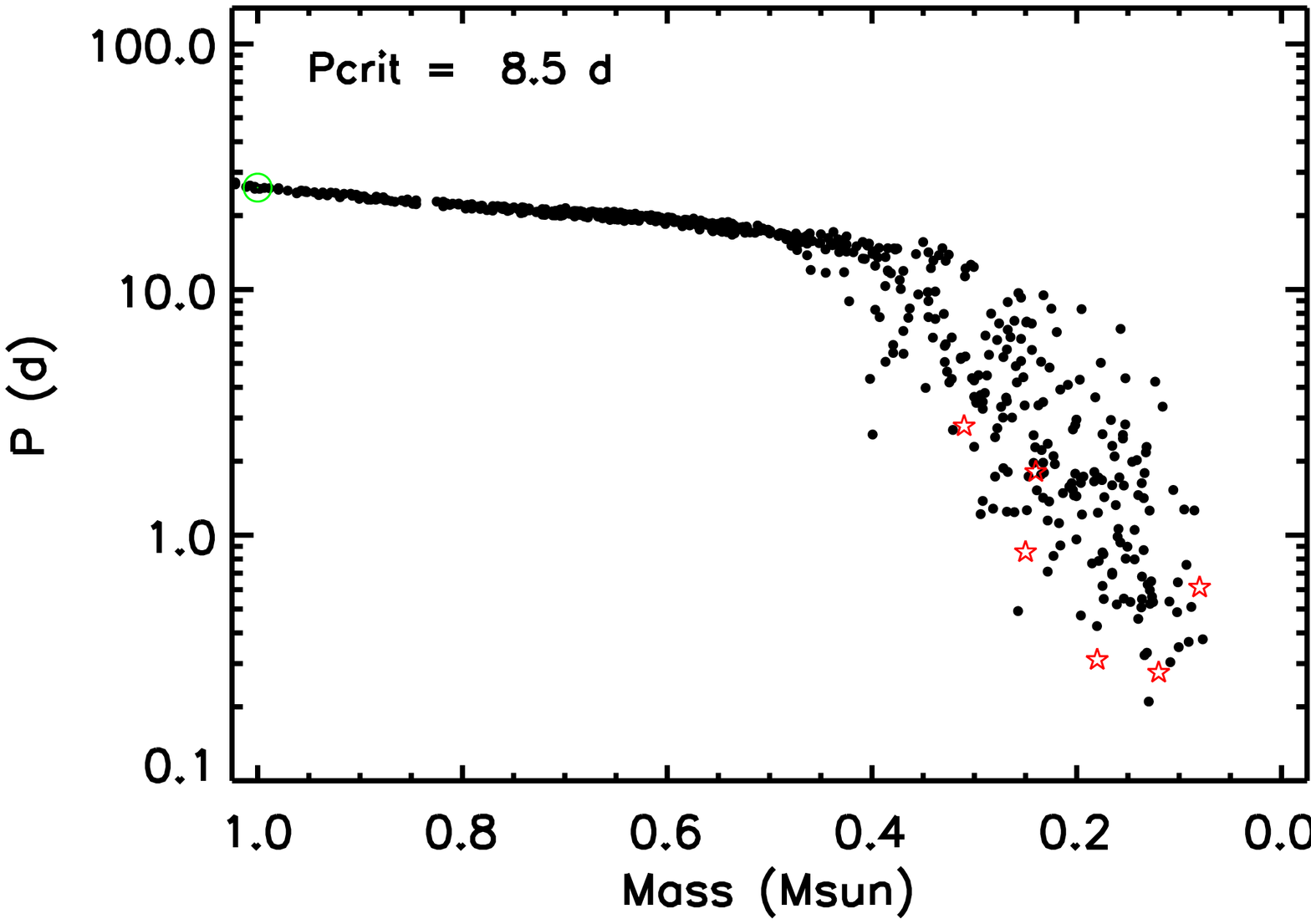}}
  \parbox{.33\textwidth}{
    \includegraphics[bb = 125 10 625 430, width=.33\textwidth,clip=]{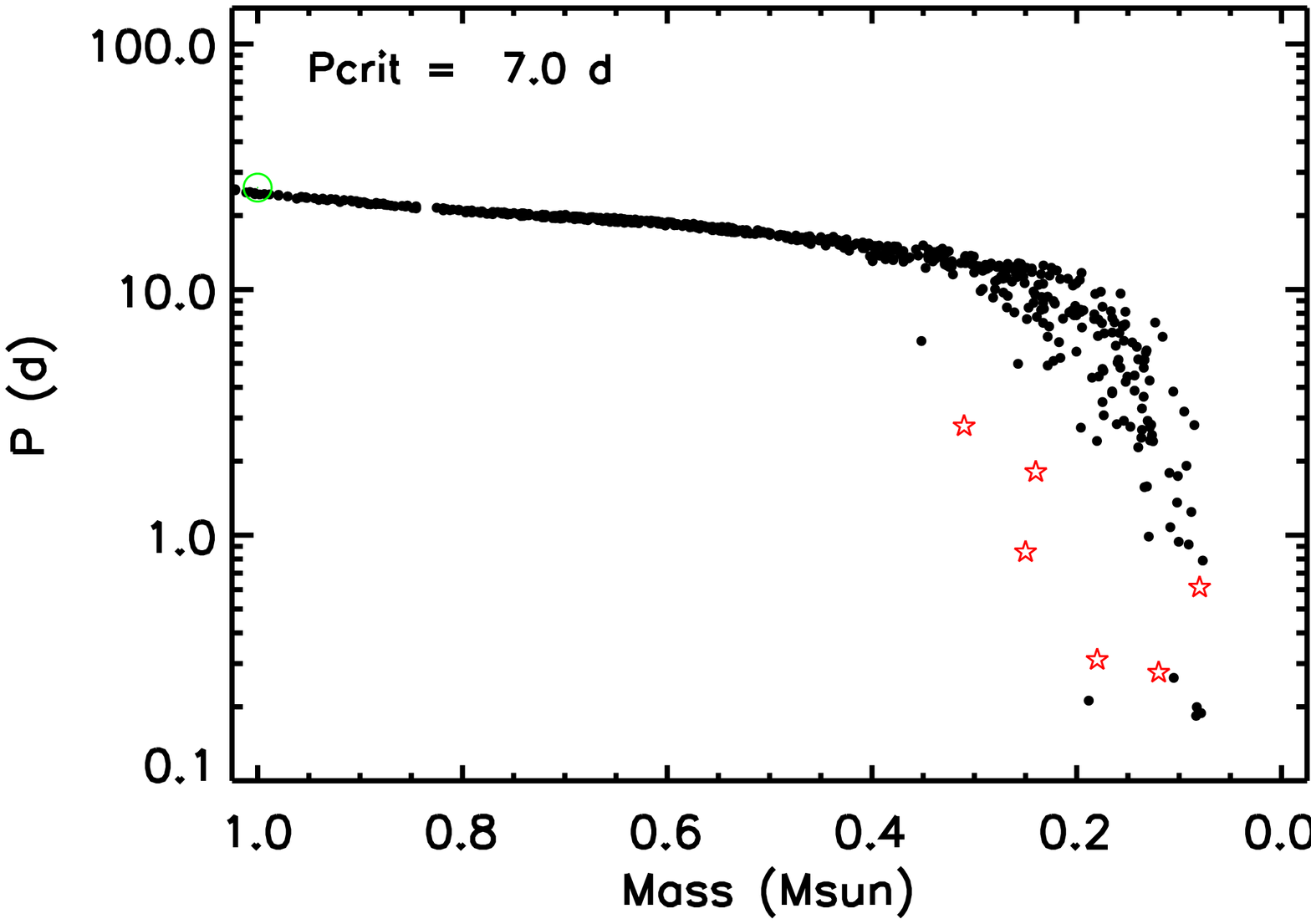}}}
\mbox{
  \parbox{.33\textwidth}{
    \includegraphics[bb = 125 10 625 430, width=.33\textwidth]{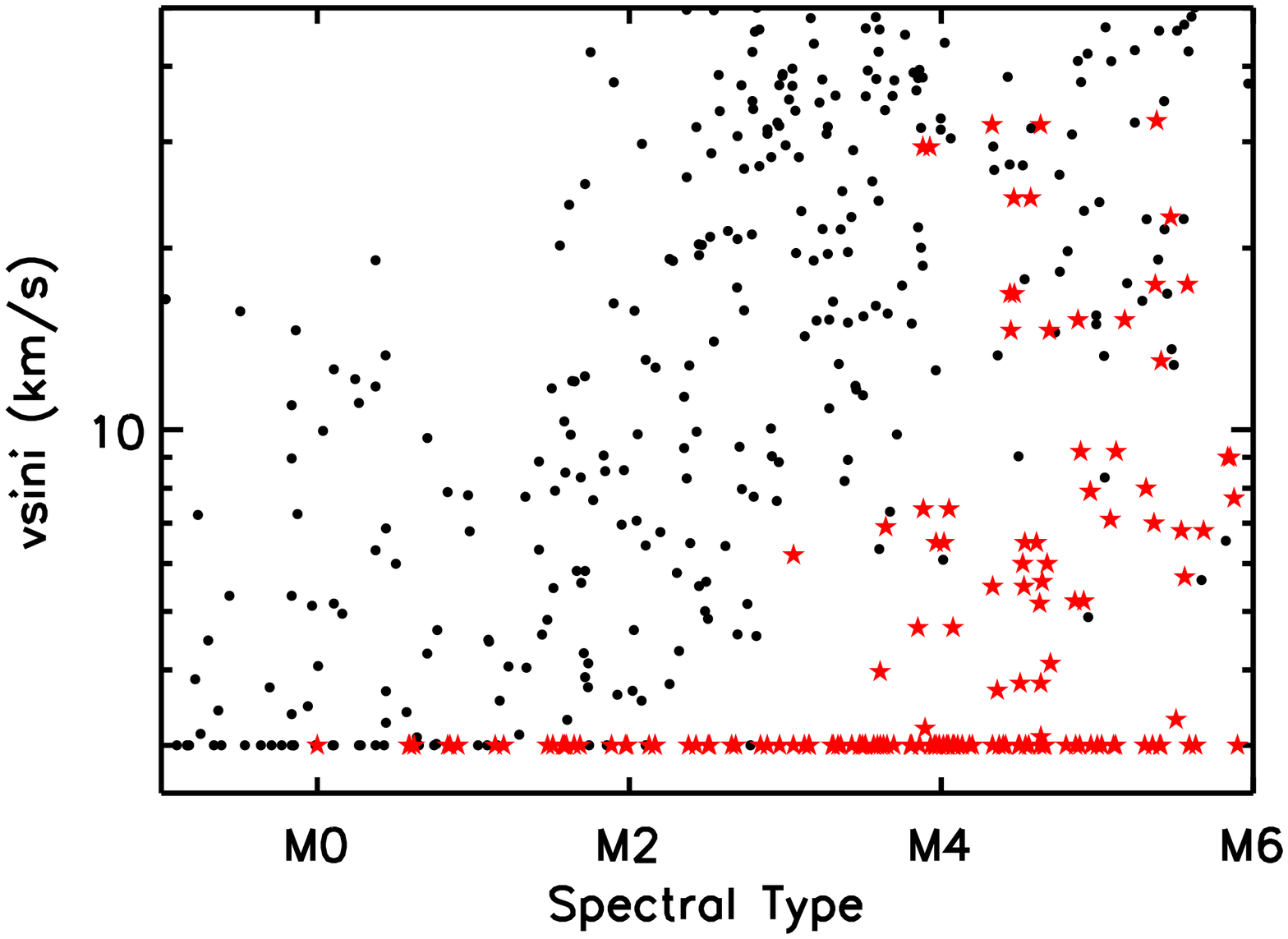}}
  \parbox{.33\textwidth}{
    \includegraphics[bb = 125 10 625 430, width=.33\textwidth,clip=]{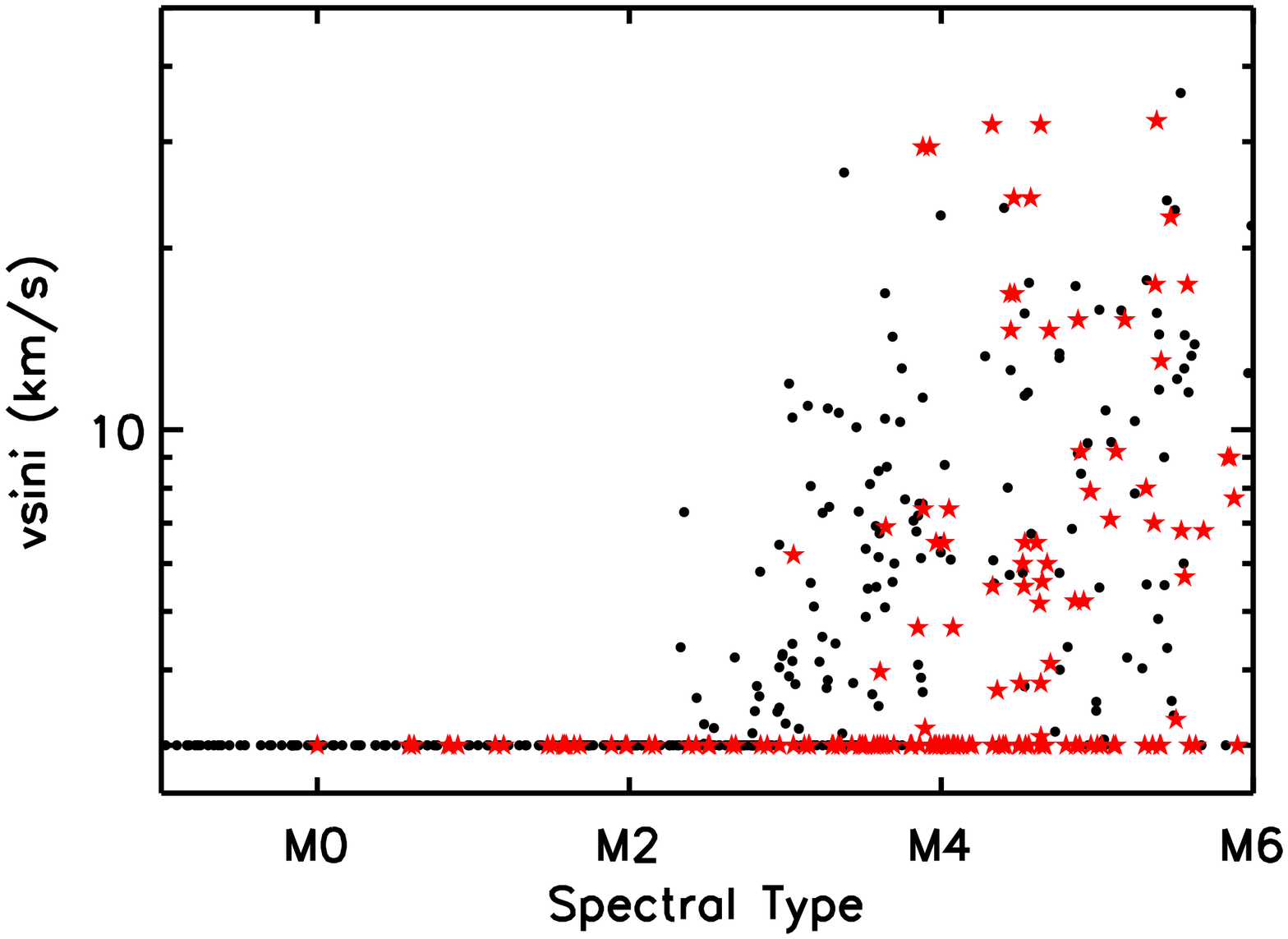}}
  \parbox{.33\textwidth}{
    \includegraphics[bb = 125 10 625 430, width=.33\textwidth,clip=]{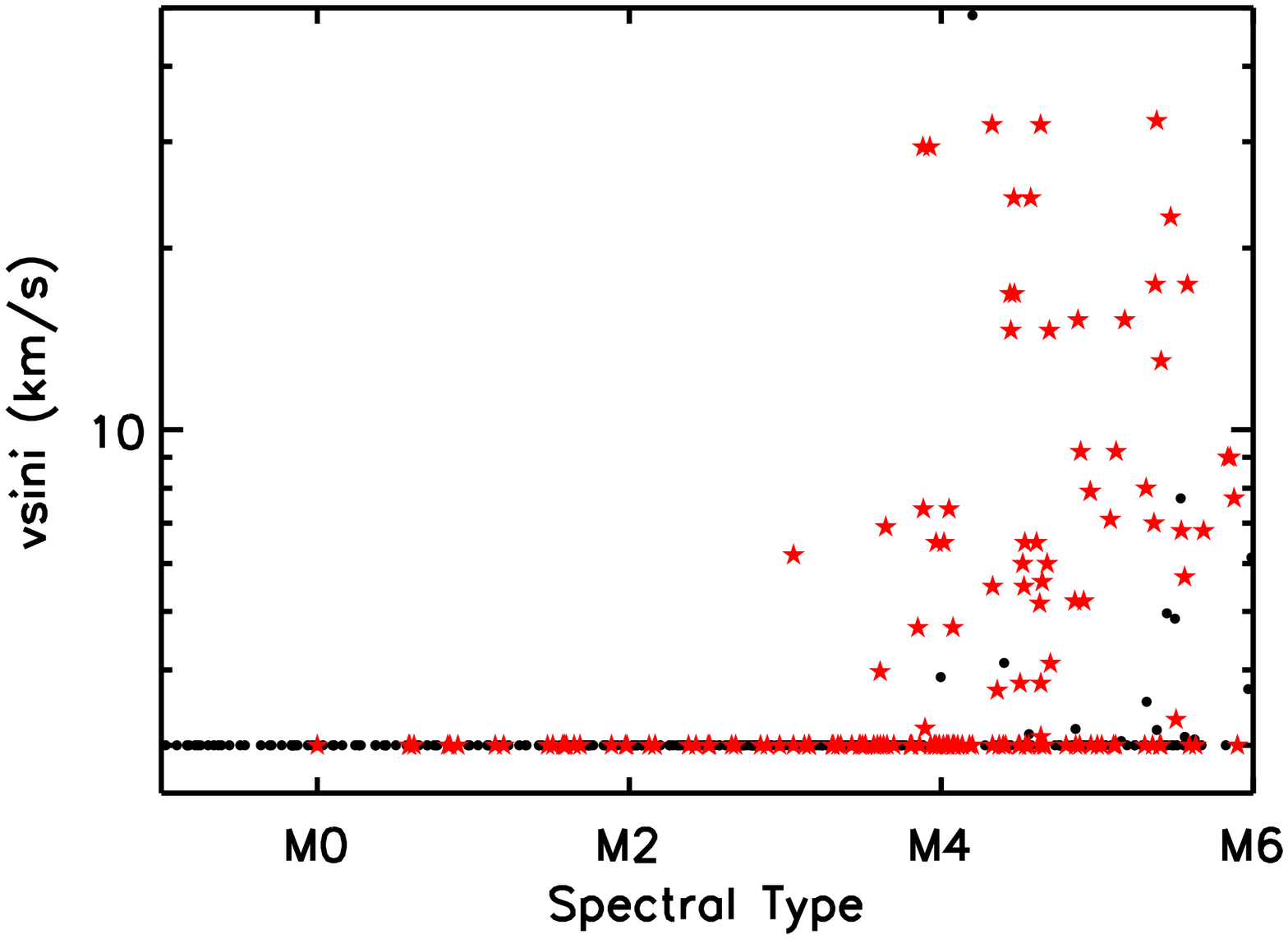}}}
\caption{\label{fig:Pcrit}Distribution of rotational periods
  (\emph{upper} panel) and surface rotation velocities (\emph{lower}
  panel) among fields stars (red stars; period data from
  \citet{2011ApJ...727...56I}; $v\,\sin{i}$ data sample explained in
  \citet{2008ApJ...684.1390R}) and according to our model at an age of
  3\,Gyr (black circles). The Sun is shown as a green circle in the
  upper panel plots. Three model calculations with different values of
  $P_{\rm crit}$ are shown; from left to right: $P_{\rm crit} = 10,
  8.5$, and 7\,d.}
\end{figure}

Fig.\,\ref{fig:Pcrit} shows the above initial distribution evolved to
an age of 3 Gyr, for various values of $\Omega_{\rm crit}$ and
$\mathcal{C}$, compared to data for the Sun and other young-disk field
low-mass stars\footnote{The age of the Sun is 4.5 Gyr; the ages of the
  field stars shown are not precisely known, but expected to lie in
  the range 1--5 Gyr (young disk).  We thus choose an evolutionary age
  of 3 Gyr for our model as a reasonable mean to compare to the Sun
  and these stars; changing this by $\sim\pm$2 Gyr has no substantial
  effect on our final results.}.

We plot the results in both period and $v\,\sin{i}$ domains, because
there are significantly more field dwarfs with $v\,\sin{i}$
measurements than with known periods.  We have converted our model
periods to velocities $v$ using the BCAH98 MS mass-radius
relationship, and $v$ to $v\,\sin{i}$ assuming $\sin{i}=\sqrt{3}/2$,
the mean value for a random distribution of inclinations.  The latter
conversion is only true in a statistically averaged sense, and not
strictly accurate for comparison to a single observed population
(which represents only one instantiation of all possible $\sin{i}$
distributions, not the average).  A mathematically rigorous comparison
between the model velocities and $v\,\sin{i}$ data requires involved
statistical analyses \citep{1993AA...269..267G}, best accomplished
with detailed Monte Carlo simulations \citep{2000MNRAS.319..457C}.  In
our case, however, the $v\,\sin{i}$ plots are only used to illustrate
more clearly the stellar mass (spectral type) at which there is a
sharp break in the rotation distribution, and to show that our
best-fit model reproduces this break both in period and velocity
space.  The average value of $\sin{i}$ is sufficient for this limited
purpose: statistical variations in the $\sin{i}$ distribution should
not significantly change the presence or location of the very steep
observed transition from a large population of undetected $v\,\sin{i}$
to a similarly large population of high $v\,\sin{i}$.

We further note that the observed stars plotted in the period-mass
panels are only those shown to belong kinematically to the young
(thin) disk population, via a careful position-dependent velocity
analysis by \citet{2011ApJ...727...56I}.  While the total sample of
field stars with known periods is significantly larger (see
compilation by \citet{2011ApJ...727...56I}), most of these do not have
kinematic ages determined as accurately.  Consequently, by including
them one risks vitiating the true young disk population with
significantly older stars, especially at longer periods (where
old-disk/halo stars dominate; see \citet{2011ApJ...727...56I}).  We
have therefore excluded these from the present analysis (for the same
reason, we have also excluded stars found by Irwin et al.\,\,to be
kinematically ``intermediate'' between the thin- and thick-disk
populations).  For stars shown in the $v\,\sin{i}$-mass panels (from
\citet{2008ApJ...684.1390R}), the kinematic age is not as
well-determined.  However, the observed break in the velocity
distribution is at a spectral type $\sim$M3 (mass $\sim$0.35
M$_{\odot}$), with a velocity detection threshold of $\sim$3
km\,s$^{-1}$.  Using the MS radii for stellar masses $\lesssim$0.35
M$_{\odot}$ from the BCAH98 tracks, one finds that the detected
velocities correspond to periods $<$5\,d, and in most cases
$\lesssim$1\,d (the fact that these are projected velocities makes the
real periods even shorter).  At such short periods, young-disk and
old-disk/halo stars appear to have a similar period-mass distribution
(see \citet{2011ApJ...727...56I}, especially their Fig.\,11), so
assuming a young-disk age should not significantly skew our results in
the $v\,\sin{i}$-mass parameter space.

Notice first that, independent of the precise choice of [$\Omega_{\rm
  crit}$, $\mathcal{C}$], the model reproduces the qualitative shape
of the data from solar-type stars down to VLMS remarkably well: slow
and nearly constant rotation periods (undetected $v\,\sin{i}$) down to
some threshold mass, followed by a sharp transition to faster rotation
with decreasing mass (later type).  This arises directly from the
strong radius-dependence of our angular momentum loss-rate, as
discussed earlier.  Such a qualitative match over the entire
0.1--1\,M$_{\odot}$ range has not been possible with previous models
based on the K88 formalism (without invoking ad hoc mass dependencies
specifically {\it constructed} to fit the data), and bolsters our
physically-motivated picture.

For a quantitative match, we simultaneously fit the position of the
Sun and the mass (or spectral type) at which the VLMS periods (or
$v\,\sin{i}$) turn over.  The plot shows models with $P_{\rm crit}$
$\equiv$ 2$\pi/\Omega_{\rm crit}$ = 7--10\,d; for each latter value,
$\mathcal{C}$ is chosen to reproduce the observed rotation period of
the Sun, yielding $\mathcal{C}$ = 4.43$\times$10$^3$ --
8.86$\times$10$^2$ (gm$^5/$cm$^{10}$s$^{3}$)$^{1/3}$.  With the Sun
fixed, the turnover in the data at spectral type $\sim$M3 (mass
$\sim$0.35 M$_{\odot}$) requires $7\,{\rm d} < P_{\rm crit} < 10\,{\rm
  d}$.  $P_{\rm crit} \ge 10$\,d cannot match the very slow rotation
(undetected $v\,\sin{i}$) at $\ge$$0.35$\,M$_{\odot}$ / earlier than
M3: the sharp break in the $v\,\sin{i}$ distribution is predicted to
occur at an earlier spectral type than observed (bottom left panel in
Fig.\,\ref{fig:Pcrit}).  Conversely, $P_{\rm crit} \le 7$\,d cannot
fit the fast rotation at $<$$0.35$\,M$_{\odot}$ / later than M3: the
sharp turnover in model rotation rates happens at a later spectral
type/lower mass than in the data (right panels of both period and
$v\,\sin{i}$ distributions in Fig.\,\ref{fig:Pcrit}).  We find that
$P_{\rm crit}$ = 8.5\,d best matches the turnover in the rotation data
(middle panels).  Our best-fit choice is thus [$\Omega_{\rm crit}$,
$\mathcal{C}$] = [8.56$\times$10$^{-6}$ s$^{-1}$, 2.66$\times$10$^{3}$
(gm$^5/$cm$^{10}$s$^{3}$)$^{1/3}$].  We defer a physical
interpretation of these values to \S\S8.2 and 9; for now, we
incorporate them into our model to test the theory at other ages.

\subsubsection{Comparisons to Open Clusters}

\begin{figure}
\mbox{
  \parbox{.33\textwidth}{
    \includegraphics[bb = 125 10 625 430, width=.33\textwidth]{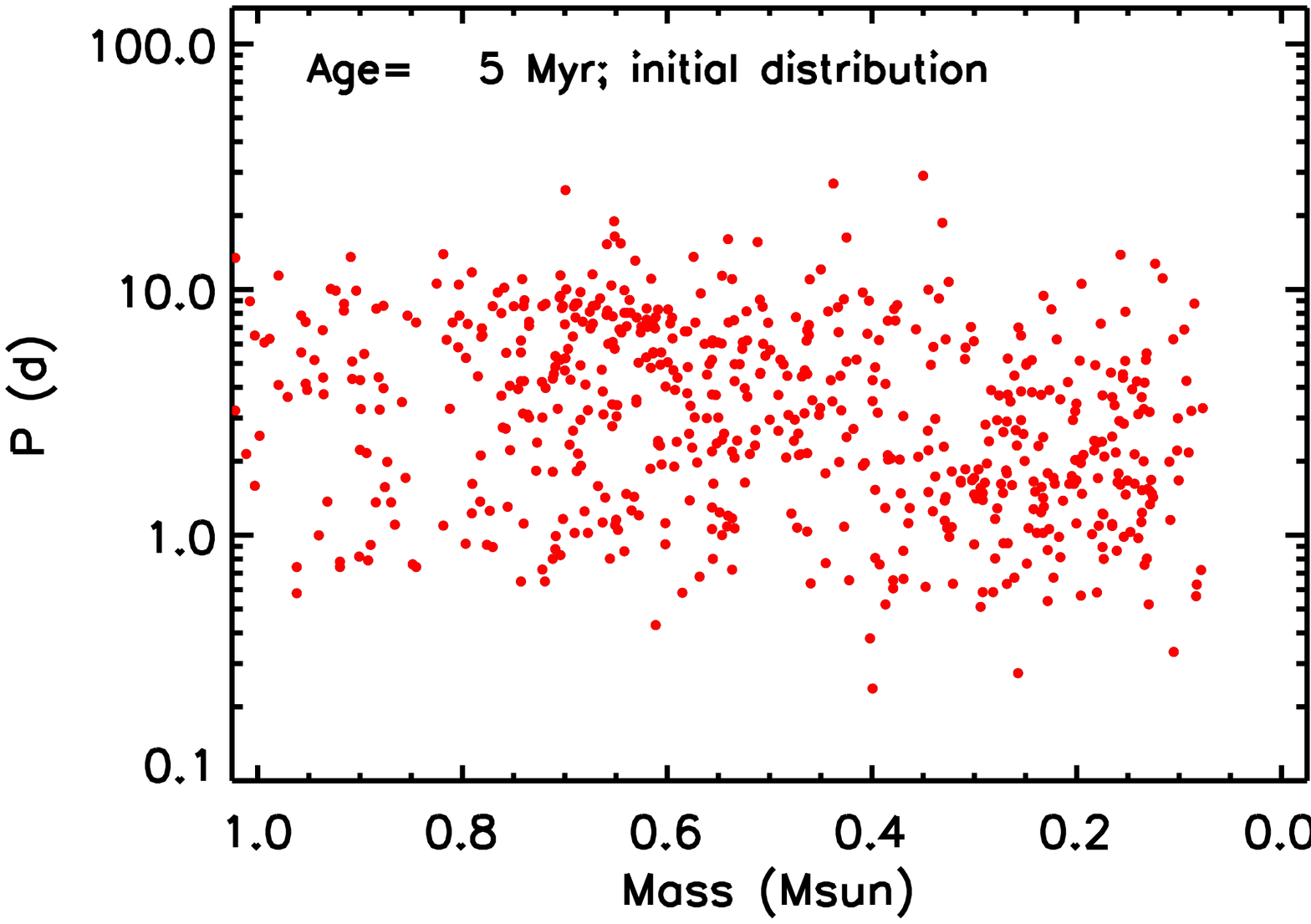}}
  \parbox{.33\textwidth}{
    \includegraphics[bb = 125 10 625 430, width=.33\textwidth,clip=]{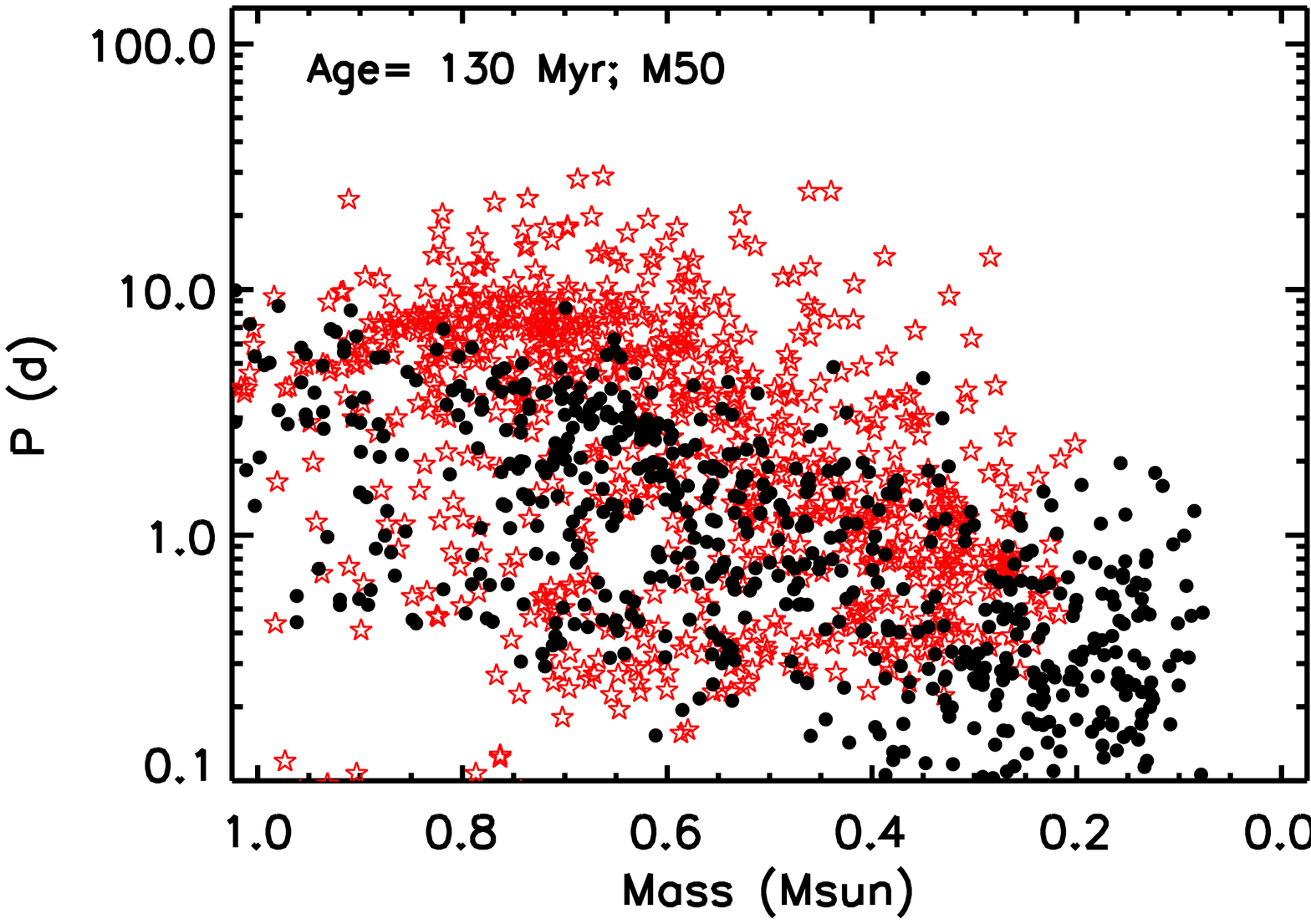}}
  \parbox{.33\textwidth}{
    \includegraphics[bb = 125 10 625 430, width=.33\textwidth,clip=]{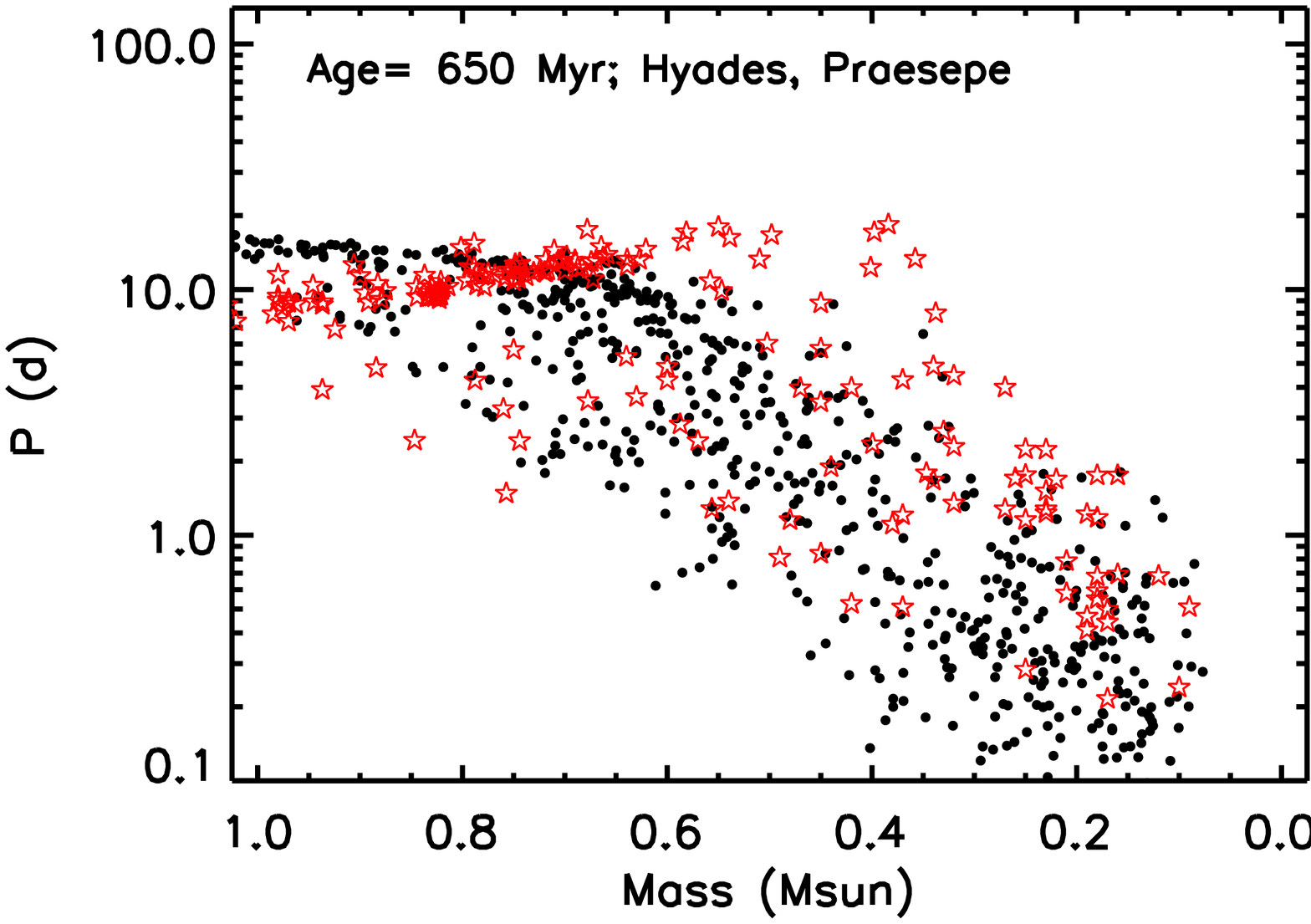}}}
\caption{\label{fig:periods}Evolution of rotational periods. The
  initial angular momentum distribution taken from ONC and NGC~2264
  (left panel, see text) is assumed. In the other two plots, we show
  angular momentum evolution of the initial sample as black points and
  observations of clusters at different ages as red stars. Data taken
  from the literature: ONC: \citet{2002A&A...396..513H}; NGC~2264:
  \citet{2005A&A...430.1005L}; M50: \citet{2009MNRAS.392.1456I};
  Hyades, Praesepe: \citet{1987ApJ...321..459R, 1995PASP..107..211P,
    2007MNRAS.381.1638S, 2011MNRAS.tmp..261S, 2011MNRAS.413.2218D}.}
\end{figure}

The results are plotted in Fig.\,\ref{fig:periods}.  The first panel is simply
the rotation data for our combined sample of ONC + NGC~2264, representing our
model distribution of rotation periods at the end of the disk locking phase at
5 Myr, as discussed earlier.  The second panel shows this distribution evolved
via our theory to an age of 130 Myr, using the best-fit [$\Omega_{\rm crit}$,
$\mathcal{C}$] inferred above.  For comparison we plot the data for the coeval
M50 open cluster.  We see that the {\it lower} envelope of data periods is
clearly inclined from $\sim$1--0.5\,M$_{\odot}$: with the exception of a few
extremely rapid rotators at $\sim$0.1\,d between $\sim$1--0.8 $M_{\odot}$, the
fastest rotation rate observed increases with decreasing mass.  This tilted
lower envelope of rapid rotators seems to be a universal feature of clusters
at this age \citep{2007ApJ...669.1167B, 2009IAUS..258..363I}.  Crucially, our
simple model quantitatively matches this envelope very well.  Note that the
shape of the envelope is {\it not} simply due to PMS spin-up: as we have
pointed out (Fig.\,2), spin-down does have some effect even during the PMS
phase, and moreover stars down to 0.5 M$_{\odot}$ have all arrived on the MS
before 130 Myr.  Instead, it arises in our model from the strong radius
dependence of the angular momentum loss-rate: less massive stars have a
smaller radius and thus a lower intrinsic braking efficiency.

Equally clearly, we do {\it not} reproduce the upper envelope of
slowest rotators for masses $\gtrsim$0.3\,M$_{\odot}$.  The most
likely reason is core-envelope decoupling, wherein only the outer
convective layer is spun down rapidly by the wind (producing the slow
{\it surface} rotation that is observed), while the inner radiative
core only spins down over longer timescales dictated by the
inefficient ``coupling'' via which it transfers angular momentum to
the outer convective layer (lengthening the overall stellar spin-down
timescale).  Specifically, it is suggestive that all low-mass stars
down to $\sim0.3$\,M$_{\odot}$ (the convective boundary, below which
stars are always fully convective) develop a radiative core by
$\sim$130\,Myr, and stars $\gtrsim 0.6$\,M$_{\odot}$ do so by
$\lesssim 50$\,Myr (see Fig.\,\ref{fig:efficiency}).  In this case,
the ``hump''-shaped upper envelope of slow rotators observed for
masses $\gtrsim$0.3 M$_{\odot}$ is precisely what core-envelope
decoupling would predict: stars that have just formed a radiative core
($\sim$0.3 M$_{\odot}$ at 130 Myr) would be starting to evince longer
periods due to decoupling; this lengthening of periods would be
maximised in more massive stars that formed a radiative core earlier
and are currently strongly decoupled (which our plot suggests occurs
around 0.4--0.6 M$_{\odot}$ at 130 Myr); and the periods would decline
again towards still more massive stars in which the time since the
formation of the core is approaching the coupling timescale.
Concurrently, stars in which no radiative core has formed yet should
evince no increase in period due to decoupling; this is indeed what
our plot shows for masses $<$0.3 M$_{\odot}$ (always fully
convective), whose upper envelope of periods (admittedly defined by
only a few observed stars in M50) agrees well with our model of
solid-body rotation.

Our simple theory, with only solid-body rotation, does not account for
core-envelope decoupling.  On the other hand, the model of
\citet{2009IAUS..258..363I}, based on the K88 formulation but
including core-envelope decoupling phenomenologically, does produce
the very long observed periods in solar type stars at $\sim$130\,Myr;
conversely, it cannot account for the fast rotation of fully
convective field stars (which should rotate as solid bodies), which
our theory does (Fig.\,3).  We thus postulate that the strongly
radius-dependent spin-down in our model (absent in K88 and
\citet{2009IAUS..258..363I}), combined with core-envelope decoupling
(absent in our theory), should yield a good fit to all low-mass stars
at 100--200 Myr (as we have qualitatively argued above for M50).  This
will be addressed in a forthcoming paper.

The third panel of Fig.\,\ref{fig:periods} shows our model evolved to
650\,Myr, compared to the combined data for Hyades and Praesepe.  The
match between model and data is now significantly better than at 130
Myr.  The model reproduces very well the {\it mean} period of the
upper envelope of 1--0.7 M$_{\odot}$ stars ($\sim$10\,d); the upper
envelope for stellar masses $\lesssim$0.3 M$_{\odot}$ (which are all
fully convective stars); and the lower envelope of periods for stars
down to 0.4 M$_{\odot}$.  All 1--0.1 M$_{\odot}$ stars have reached
the MS and thus stopped spinning up by $\sim$600 Myr, so these good
fits are all strongly linked to the radius-dependent angular momentum
loss in our theory.

What we do not reproduce is the gently rising upper envelope of
periods with decreasing mass down to 0.4 M$_{\odot}$ (our theory
predicts a declining upper envelope with mass over the entire 1--0.1
M$_{\odot}$ range), and the upper envelope in general from 0.7 to 0.3
M$_{\odot}$.  These discrepancies, and the overall convex shape of the
upper envelope of periods from 1 to 0.3 M$_{\odot}$, are again almost
certainly due to core-envelope decoupling.  As described earlier,
stars that have just formed a radiative core should just be starting
to exhibit the longer periods associated with decoupling (essentially
no such stars at 650 Myr: see end of this paragraph); more massive
stars with cores formed earlier should be strongly decoupled (our plot
indicates this occurs around 0.4 M$_{\odot}$ at 650 Myr); and even
more massive stars, in which the time since core formation is becoming
comparable to the coupling timescale, should evince a decline in
periods with increasing mass due to burgeoning coupling, as is
observed.  The veracity of this scenario is bolstered by three
additional trends.  First, the peak of the upper envelope is clearly
shifted to lower masses going from 130 to 650 Myr (from a plateau over
$\sim$0.6--0.4 M$_{\odot}$ in M50 to a peak at 0.4 M$_{\odot}$ in
Hyades/Praesepe), which is what core-envelope decoupling predicts
(since lower mass stars form radiative cores later, they are strongly
decoupled later as well).  Second, stars with mass $\lesssim$0.3
M$_{\odot}$ remain fully convective even on the MS, and thus should
not exhibit any decoupling effects.  This is indeed what we see: the
upper envelope at these masses is an excellent match to our model of
solid-body rotation.  Third, we see a steep increase in periods
towards stars slightly more massive than the fully convective boundary
at $\sim$0.3 M$_{\odot}$, with the peak in the period distribution
already reached by 0.4 M$_{\odot}$.  This is also explained by
core-envelope decoupling: since all stars down to the convective
boundary have developed a radiative core by $\sim$130 Myr, i.e., well
before 650 Myr, stars slightly more massive than this boundary are all
already strongly decoupled at 650 Myr.  Thus, at this age, there
should indeed be a sharp rise in periods from fully convective (fully
coupled) stars to those slightly more massive with a radiative core
(fully decoupled).

In this picture, the good agreement noted earlier, between the data
and our model in the mean period of the upper envelope of 1--0.7
M$_{\odot}$ stars, indicates that these stars are nearly fully coupled
again.  The implied coupling timescale (time elapsed between core
formation in these stars, at $<$\,50 Myr, and the onset of good
coupling) is thus $<$600 Myr, as also found by
\citet{2009IAUS..258..363I} through an explicit modeling of
core-envelope decoupling.

As an aside, we note that the masses \citet{2011MNRAS.413.2218D}
derive for their Hyades sample imply periods of 10--20\,d for some
stars down to 0.2 M$_{\odot}$, considerably slower than the upper
envelope of more massive solar-type stars (see their Fig.\,15).
However, the upper envelope of initial periods in SFRs is relatively
flat with mass at 10--20\,d (e.g., our ONC + NGC 2264 sample;
specifically, there is no evidence of slower initial periods or more
efficient disk-locking in VLMS; if anything, the opposite is more
likely).  Moreover, stars $\lesssim$\,0.3 M$_{\odot}$ cannot undergo
core-envelope decoupling (they are always fully convective).  It is
thus very hard to understand how these stars can rotate much slower on
the early MS than solar-type stars, which arrive on the MS much
earlier and thus have a far shorter PMS spin-up phase, and which
undergo core-envelope decoupling to boot.  However, accounting for the
scatter in mass introduced by Delorme et al.'s ($V-K$) color-dependent
mass-determination technique, and correcting for this with our
absolute magnitude-based method (see \S6), we find that these large
periods are actually associated with masses $>$0.3 M$_{\odot}$, i.e.,
stars {\it more massive} than the fully convective boundary (as shown
in our plot).  This removes the dilemma, since core-envelope
decoupling can now fully explain the observed large periods, as
described above, and maximum periods are now shorter in fully
convective stars than in solar-types, as expected.  This undercores
the need for good mass estimates for understanding rotation evolution.

In summary: Our theory predicts rotation in low-mass stars to be
fundamentally radius- and hence mass-dependent.  Fixing our two free
model parameters, $\Omega_{\rm crit}$ and $\mathcal{C}$, via
comparison to the Sun and other roughly coeval field stars, enables us
to quantitatively reproduce many of the features of the observed
mass-rotation distribution from 130 Myr up to a few Gyr
(Figs.\,\ref{fig:Pcrit}, \ref{fig:periods}), without invoking
variations in dynamo mode or field topology.  The trends that we do
{\it not} replicate are all qualitatively explicable with the addition
of transient core-envelope decoupling (not included in our model),
which is important for ages intermediate between the formation of a
radiative core and the resumption of good coupling $<$600 Myr later.
This will be quantitatively verified in our next paper.  After the
decoupling phase in solar-type stars, and always for fully convective
VLMS, our radius-dependent theory is in excellent agreement with the
data up to a few Gyr.

\subsection{Mass (Rossby Number) Dependence of $\Omega_{\rm crit}$}

\begin{figure}
\mbox{
  \parbox{.5\textwidth}{
    \includegraphics[width=.48\textwidth,clip=]{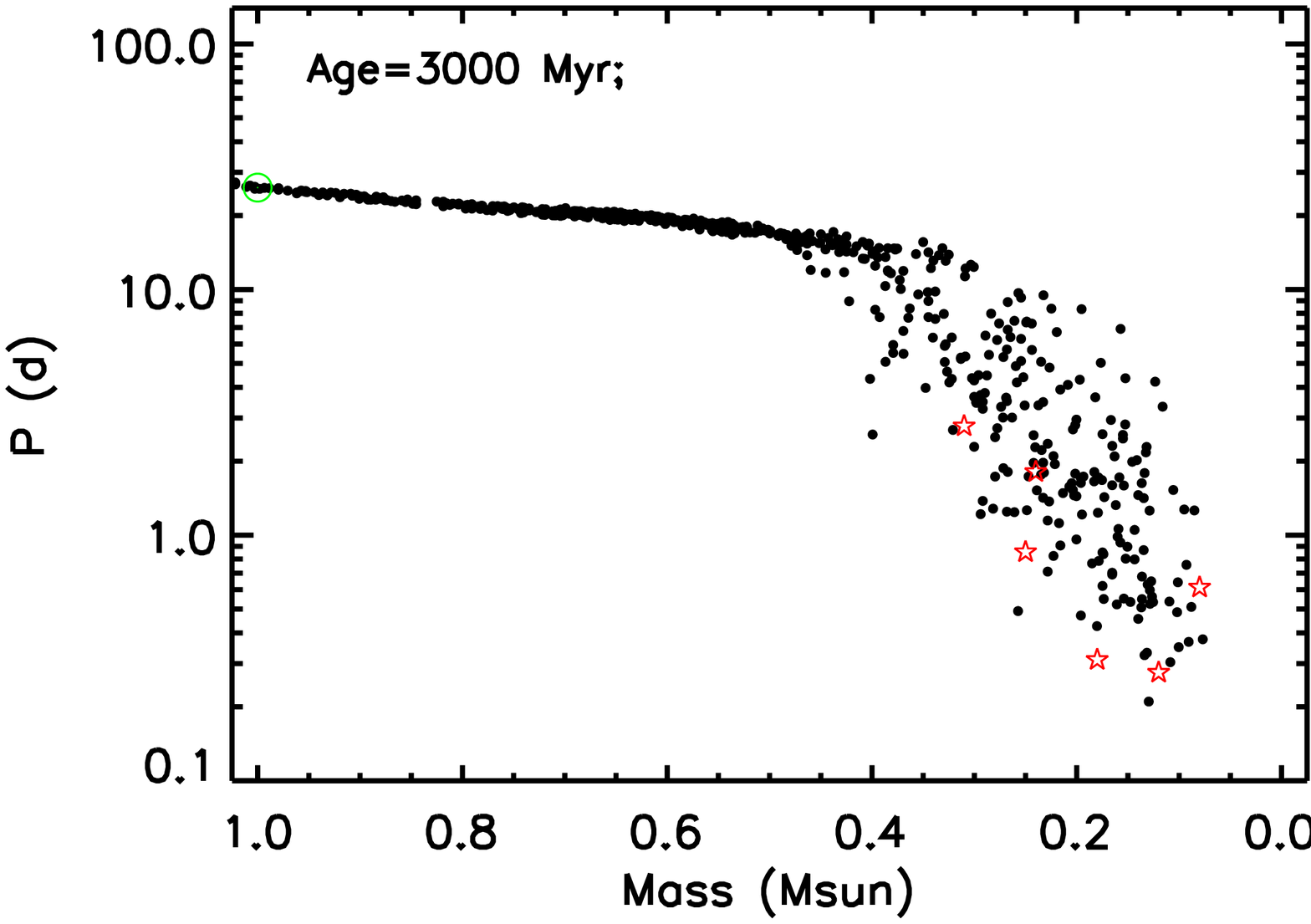}}
  \parbox{.5\textwidth}{
    \includegraphics[width=.48\textwidth,clip=]{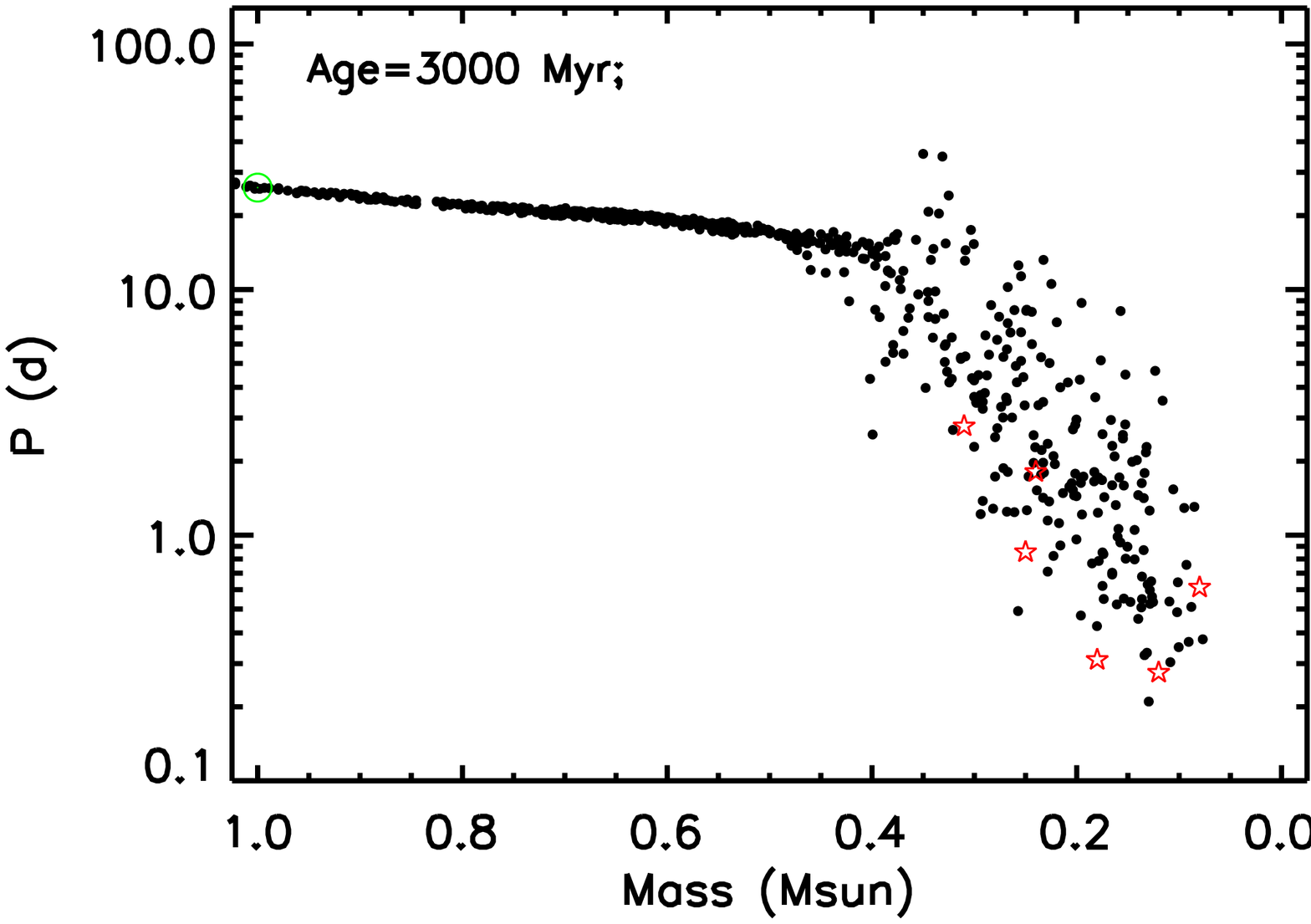}}}

\mbox{
  \parbox{.5\textwidth}{
    \includegraphics[width=.48\textwidth,clip=]{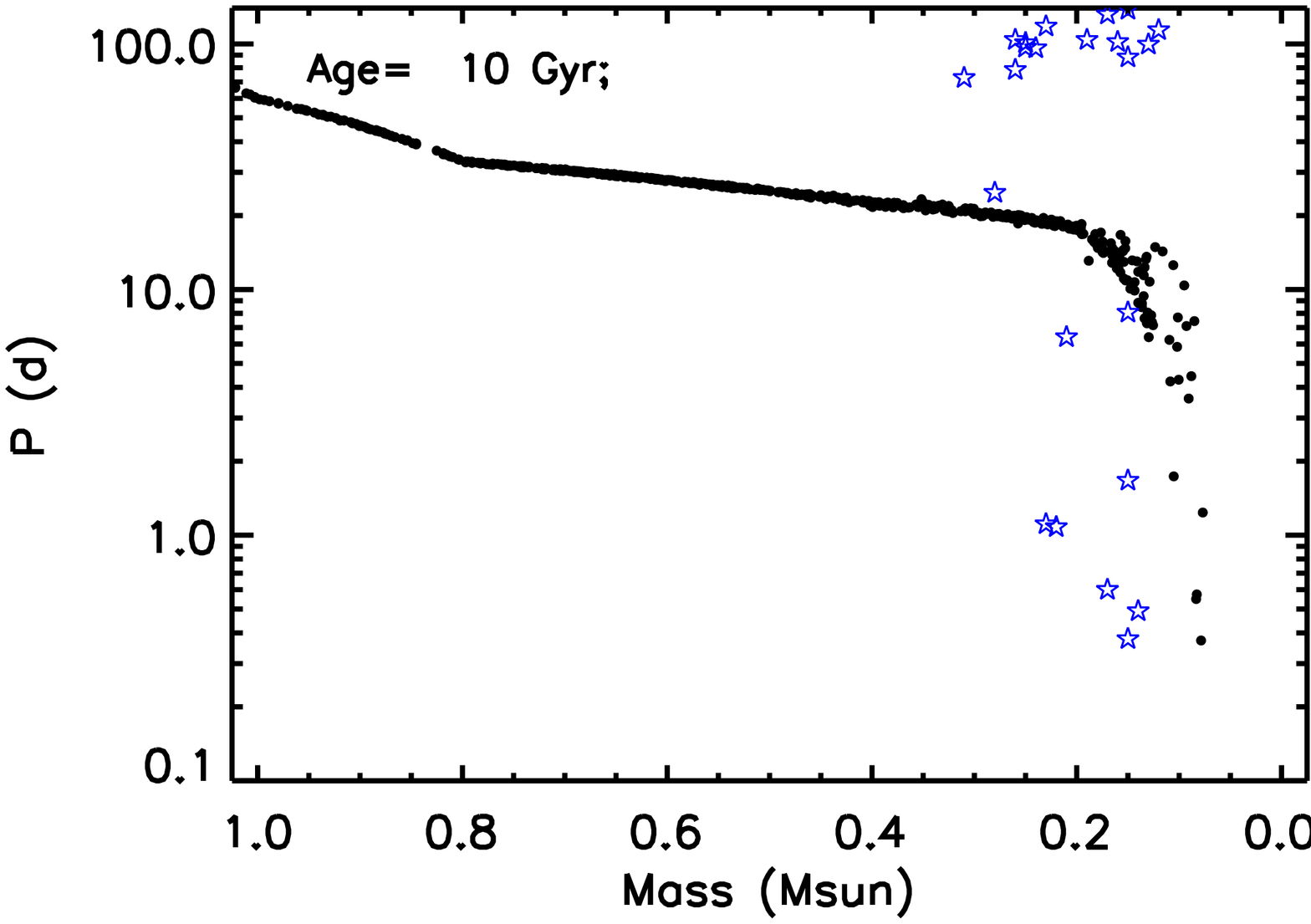}}
  \parbox{.5\textwidth}{
    \includegraphics[width=.48\textwidth,clip=]{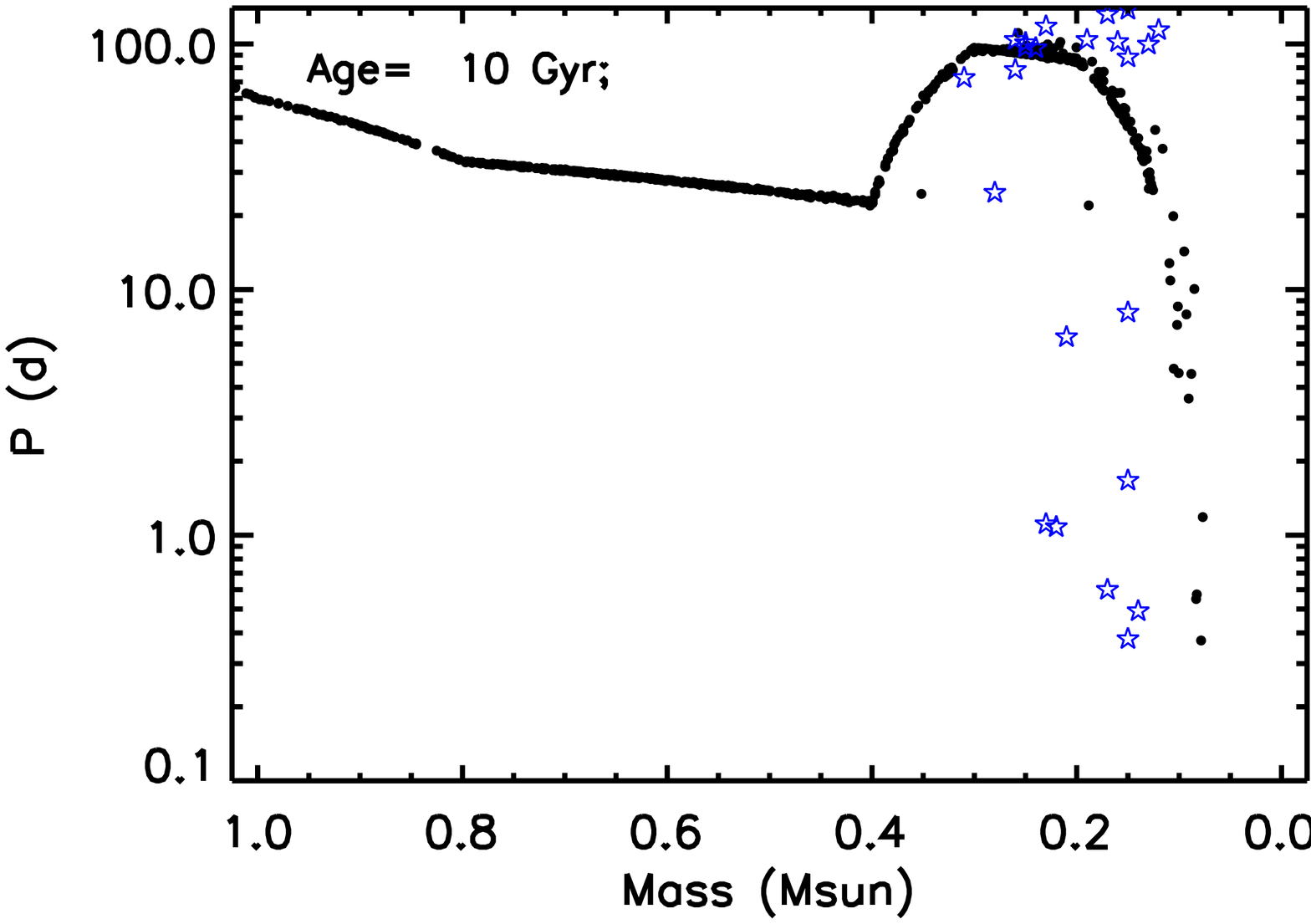}}}
\caption{\label{fig:P10}Model distribution of rotational periods at
  ages 3\,Gyr (upper panel) and 10\,Gyr (lower panel). \emph{Left
    panel:} Model with one critical rotation period $P_{\rm crit} =
  8.5$\,d for all stars (upper left panel is identical to top middle
  panel in Fig.\,\ref{fig:Pcrit}). \emph{Right panel:} Model using
  $P_{\rm crit} = 8.5$\,d for stars with $M > 0.4$\,M$_{\odot}$ and
  $P_{\rm crit} = 40$\,d for less massive stars. Blue and red stars
  show measurements of rotation periods \citep{2011ApJ...727...56I} in
  M stars, red stars are young disc objects, blue stars are halo
  objects.}
\end{figure}

There is still a last wrinkle.  Fig.\,\ref{fig:P10} (bottom left)
shows our model evolved to 10 Gyr, compared to data for old-disk/halo
field M dwarfs recently published by \citet{2011ApJ...727...56I} (ages
$\sim$7--13 Gyr).  Most of the observed periods at
$\lesssim0.3$\,M$_{\odot}$ are very long -- 20 to $>$100\,d -- while
our model barely reaches 20\,d at these masses\footnote{The spin-down
  of unsaturated $\sim$1 $M_{\odot}$ stars to 60\,d by 10 Gyr is due
  to their {\it increase} in radius as they move {\it off} the MS by
  $\sim$3 Gyr; the same cannot apply to the VLMS, whose MS lifetimes
  exceed a Hubble time.}.  With our choice of $P_{\rm crit}$ (=8.5\,d)
and $\mathcal{C}$, slowing to observed periods an order of magnitude
longer than $P_{\rm crit}$ requires $> 10^{12}$ yr.  One solution is
to fiddle with the spin-down timescales $t_{S}$ and $t_{U}$ via
$\mathcal{C}$, or invoke radically different fields, wind velocities
etc.  Very precise fine-tuning would then be needed, however, to avoid
doing violence to the match already obtained to both solar-types and
VLMS at earlier ages; that our simple theory (complemented by
core-envelope decoupling at open cluster ages) yields this good match
argues against such physically unmotivated `fitting'.

Instead, the simplest solution is that $P_{\rm crit}$ is much larger
in stars $\lesssim0.3$\,M$_{\odot}$ (i.e., they remain saturated at
much slower rotation rates than higher mass stars).  There is good
reason to believe so, as discussed below.  For now, note that at a
young disk age of $\sim$3 Gyr, the observed stars at
$\lesssim0.3$\,M$_{\odot}$ are mostly saturated.  Thus, while they set
a lower limit on our best-fit $P_{\rm crit}$, they are insensitive to
the upper limit, which is set instead by unsaturated slow rotators
(undetected $v\,\sin{i}$) at $> 0.3$\,M$_{\odot}$ (including the Sun;
see discussion of Fig.\,\ref{fig:Pcrit} in \S8.1.1).  Hence invoking a
much larger $P_{\rm crit}$ {\it only} for stars
$\lesssim0.3$\,M$_{\odot}$ should preserve all our results upto a few
Gyrs, while enabling these stars alone to remain saturated -- and thus
spin down exponentially -- for much longer, thereby achieving far
longer periods by 10\,Gyr.  Fig.\,\ref{fig:P10} (right panels)
illustrates this, for a fiducial $P_{\rm crit}$ = 40\,d (motivated
below) for $\leq0.3$\,M$_{\odot}$; $P_{\rm crit}$ is held fixed at
8.5\,d for higher masses and $\mathcal{C}$ is unchanged for all
masses.  We see that the match to data at 3\,Gyr continues to be
excellent, while at 10\,Gyr our model now fits the very slowly
rotating old-disk/halo stars as well.

As an aside, we note the additional presence of relatively rapid
rotators around 0.2 M$_{\odot}$, offset from the tail of rapid
rotators at $\sim$0.1 M$_{\odot}$ in our 10 Gyr model.  However, these
do not represent any fundamental puzzle.  The observed stars span ages
of 7--13 Gyr, while our model is for a unique age of 10 Gyr;
implementing the observed age range in our model should allow us to
simultaneously fit the extremely slow and relatively rapid rotators
(e.g., a 7 Gyr model would evince a tail of rapid rotators at a higher
mass, more in line with the data).  We do not attempt this in our
present exploratory analysis, where we simply seek to understand the
additional physics implied by the very slow rotators; fitting the
entire period distribution of these old stars is undertaken in our
next paper.

What is the physical basis for an increased $P_{\rm crit}$ at the
lowest masses?  In dynamo theory (both $\alpha\Omega$ and $\alpha^2$;
\citet{2006A&A...446.1027C}), the magnetic field strength is
determined not by the rotation rate alone, but its ratio to the
convective turnover timescale $\tau_c$~, i.e., by the Rossby number:
$\mathcal{R} \equiv P/\tau_c$.  Saturation sets in when $\mathcal{R}$
decreases below some threshold value $\mathcal{R}_{\rm crit}$.  In
this paradigm, our $P_{\rm crit}$ is really to be interpreted as
$P_{\rm crit} \equiv \tau_c \mathcal{R}_{\rm crit}$.  There is some
empirical evidence that $\mathcal{R}_{\rm crit} \sim 0.1$
\citep{2009ApJ...692..538R}.  If we assume the latter, then our
$P_{\rm crit}$ = 8.5\,d inferred for stars spanning
1\,--\,$>$0.3\,M$_{\odot}$ implies $\tau_c \sim$ 85\,d for these
masses.  While $\tau_c$ is only well--defined within the idealized
mixing-length theory (MLT), and even then hard to characterize,
approximate MLT models indicate MS values of $\sim$40--150\,d for
1--0.5\,M$_{\odot}$ \citep{1996ApJ...457..340K}.  It is suggestive
that the mean is then indeed very close to our implied 85\,d, and the
range within a factor of 2 of this value.  At the same time,
extrapolation of the MLT models indicates $\tau_c \gtrsim 250$\,d on
the MS for masses $\lesssim 0.3$\,M$_{\odot}$.

From an empirical standpoint, the overall situation is similar, but
differs from the above estimates of $\tau_c$ in an important respect.
Using a Rossby number formalism, and examining various markers of
magnetically driven activity, \citet{1994AA...292..191S} and
\citet{2007AcA....57..149K} have investigated the convective turnover
time for $\sim$0.2--1.2 M$_{\odot}$ stars.  While their analysis does
not yield the absolute value of $\tau_c$ \citep{1994AA...292..191S},
they find that the {\it relative} (i.e., scaled) value of $\tau_c$
increases from 1.2 to 0.8 M$_{\odot}$, then {\it levels off} until
$\sim$0.5 M$_{\odot}$, and then increases steeply again till $\sim$0.2
M$_{\odot}$.  The theoretical estimates by \citet{1996ApJ...457..340K}
miss this intermediate plateau in turnover timescales: they predict a
factor of $\sim$4 increase in $\tau_c$ from 1 to 0.5 M$_{\odot}$,
while the empirical results imply a very small increase of only a
factor of $\sim$1.5.  On the other hand, both the empirical analysis
and (extrapolated) theory indicate a large increase in $\tau_c$ going
from 1 M$_{\odot}$ to VLMS at $\lesssim$\,0.3 M$_{\odot}$.  Given the
difficulties in calculating $\tau_c$ from first principles -- the
convective velocities and lengthscales are not theoretically well
determined (nor unique with depth), and the assumption of MLT
introduces further uncertainties \citep{1996ApJ...457..340K} -- the
empirical estimates of $\tau_c$ appear a better guide at present
(where $\tau_c$ is to be regarded as an ``effective'' overturn
timescale that is meaningful to the star, rather than a quantity
defined only within MLT).

Under the circumstances, it is highly suggestive that the empirical
$\tau_c$ are roughly constant for solar-type stars down to $\sim$0.5
M$_{\odot}$, and then rise sharply towards lower masses; this is
precisely the trend we have advocated above to explain the very slowly
rotating old VLMS.  To quantify this agreement, we scale the relative
$\tau_c$ values from \citet{2007AcA....57..149K} such that the mean
$\tau_c$ over 1--0.5 M$_{\odot}$ equals our best-fit value of 85\,d
for these stars (as derived above using $\mathcal{R}_{\rm crit}=0.1$;
also equal to the {\it mean} theoretical $\tau_c$ for this mass range,
as noted earlier).  We find that the effective $\tau_c$ implied by the
results of \citet{2007AcA....57..149K}, for stars $\lesssim$\,0.3
M$_{\odot}$, is then $\gtrsim$\,300\,d (close to the extrapolated
theoretical $\tau_c$ for these masses).  With $\mathcal{R}_{\rm crit}
= 0.1$, this implies $P_{\rm crit} \gtrsim 30$\,d, completely
consistent with the fiducial $P_{\rm crit}=40$\,d we have used to fit
the observed periods of these stars.  It thus appears that the
effective $\tau_c$ is indeed a physically important parameter for
angular momentum evolution.  In our next paper, we include the
observed mass-dependence of this parameter in a smoother fashion,
instead of as the step-function adopted here.

We emphasize that while we are led to a lengthening of $P_{\rm crit}$
at roughly the mass boundary for full-convection to explain the oldest
VLMS, this is a {\it separate} effect from the radius dependence of
angular momentum loss that yields a sharp turnover in periods near
this boundary at a few Gyr, and critically shapes the entire
mass-period relationship at all ages.

We further reiterate that a mass-dependent $P_{\rm crit}$ {\it alone}
cannot explain the evolution of the mass-rotation relationship; the
separate radius-dependence of $dJ/dt$ is essential.  Without the
latter, \citet{2000ApJ...534..335S} (using the K88 formalism) were
forced to conclude that a simple physically motivated mass-dependent
Rossby number scaling cannot explain rotation from solar masses to
VLMS; allied {\it with} radius-dependence, we have shown that it {\it
  can}.  The fact that the two distinct effects both have a strong
influence at masses near the fully-convective boundary is
unsurprising: as stellar masses decrease towards this boundary, both
the MS radius and luminosity decline rapidly; the former drives the
strong radius-dependence of $dJ/dt$, while the latter yields the rapid
increase in $\tau_c$ (since slower convective velocities can transport
the luminosity outwards) and hence in $P_{\rm crit}$.

\subsection{Activity Lifetimes}

Independent information on the timescales of rotational braking comes
from activity measurements. It is well established that chromospheric
and coronal emission scale with rotation in the sense that, below a
critical rotation velocity, emission is stronger with faster rotation,
while above that velocity, activity is saturated
\citep[e.g.][]{2003A&A...397..147P}. Because braking is significantly
weaker in low-mass stars, this immediately leads to the conclusion
that activity lifetimes must be significantly longer at very low
masses. Activity lifetimes of M dwarfs were determined by
\citet{2008AJ....135..785W}, who define ``lifetime'' as the typical
timescale over which H$\alpha$ can be observed in emission in their
sample, before the emission falls below their detection
limit. \citet{2008AJ....135..785W} present activity lifetimes for
spectral type bins M0--M7; we have converted the latter to stellar
masses as described in \S6.

\begin{figure}
    \includegraphics[width=16.6cm,clip=]{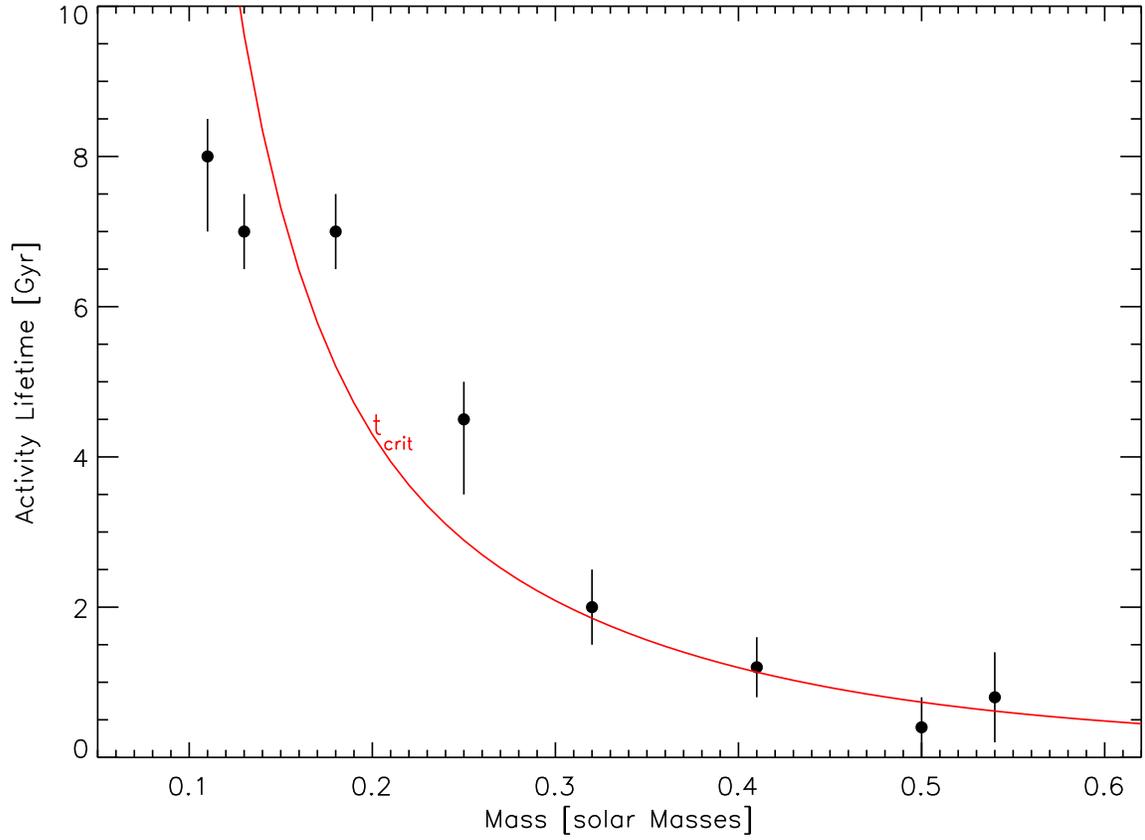}
    \caption{\label{fig:ActLifetime}Activity lifetimes of M dwarfs
      from \cite{2008AJ....135..785W} (filled circles), compared to
      the critical timescale for rotational braking in our model,
      $t_{\rm crit} = t_0 + t_S \ln{[\Omega_0/\Omega_{\rm crit}]}$
      (red line). }
\end{figure}

We compare these empirical lifetimes to the $t_{\rm crit}$ implied by
our model as a function of stellar mass, where $t_{\rm crit}$ is the
age at which a star spins down from the saturated to unsaturated field
strength regime (\S4).  The results are plotted in
Fig.\,\ref{fig:ActLifetime}. We note from the outset that there are
some caveats regarding the validity of this comparison.  First, it
implicitly assumes that saturation of the field strength and of
activity are related phenomena.  While saturation of the magnetic
field is commonly assumed to be the basis for saturated activity, the
relationship between the two must be more firmly established by future
observations.  Second, our value of $t_{\rm crit}$ is not necessarily
identical to the time at which H$\alpha$ emission becomes
undetectable, because activity diminishes rather slowly when the star
is braked in the unsaturated regime.  Third, the comparison assumes
that field strength alone determines activity levels.  For the latest
type M dwarfs, this link becomes weaker, because the low atmospheric
ionization (due to the low $T_{eff}$) means that the generation of
magnetic stresses, and hence activity, becomes inefficient even if the
field is strong \citep{2002ApJ...571..469M}.  Finally, as always,
there are inaccuracies in converting spectral types to stellar masses
via evolutionary models; these uncertainties are probably largest in
the coolest dwarfs, where mass drops very steeply with spectral type
(so small errors in the latter produce larger scatter in mass).

In spite of these caveats, Fig.\,\ref{fig:ActLifetime} shows that the
timescale for saturated braking, $t_{\rm crit}$, reproduces very well
the activity lifetimes of M stars with masses between 0.2 and
0.6\,M$_\odot$. Below 0.2 M$_{\odot}$, activity lifetimes are shorter
than $t_{\rm crit}$, but the discrepancy may be explained by the
inefficient generation of magnetic stresses due to low photospheric
ionization, as mentioned above. in summary, over a large range in
stellar mass, the activity lifetimes of chromospheric emission can be
explained by the spin-down timescales alone, without requiring a
change in the magnetic dynamo from solar-like to fully convective
stars.

\section{Discussion of Magnetic Topology and $\mathcal{C}$} 

In this work, we have assumed that the stellar fields are radial, as
the simplest possible choice.  Real surface fields, however, appear to
be a complex mixture of multipoles.  With our simple model, we have
not found any need to invoke variations in the field topology with
stellar mass to explain the data; nevertheless, is it possible that in
real stars, such variations play a role in sculpting the observed
mass-rotation distribution?

To answer this, consider the current data concerning field structure.
The basic result so far is that stars below the convective boundary
appear to have more dipolar fields, while higher-mass solar-type stars
(i.e., with radiative cores) seem to harbour a preponderance of
higher-order multipolar fields \citep{2010MNRAS.407.2269M}.  Prima
facie, however, this change goes in the {\it wrong direction} to
explain the observed trend in rotation periods: fields ordered on
larger scales (e.g., dipoles) should lead to higher rates of angular
momentum loss than fields ordered on smaller scales (i.e.,
higher-order multipoles).  This would lead to {\it slower rotation in
  fully convective stars compared to solar-types}, not higher as
observed.

Additionally, the change in field topology in fully convective stars
does not appear monotonic; stars later that 0.2 M$_{\odot}$ appear to
become less dipolar again, similar to solar-types and unlike fully
convective stars with mass $\gtrsim$0.2 M$_{\odot}$
\citep{2010MNRAS.407.2269M}. So a change in magnetic topology cannot
even be invoked in the same way for all fully convective stars, apart
from the serious problem with the expected trend discussed above.
Fundamentally, we believe that substantially more data, and a more
careful examination of the selection effects for the stars with
measured field structure, is required before any firm conclusions can
be drawn about how field topology actually changes from solar-type to
fully convective objects.

This does not mean, however, that the field structure is unimportant
for angular momentum evolution. One possible effect of the field
structure becomes clearer upon considering the constant $\mathcal{C}$
in our model.  For our best-fit value of $\mathcal{C}$, and making the
standard assumption $K_V$=1, we find that: {\it (a)} if we assume the
standard solar value for $\mdot$ = $10^{-14}$ M$_{\odot}$\,yr$^{-1}$,
then $B_{\rm crit}$ = 20 G, which is far too small; and {\it (b)}
conversely, if we assume $B_{\rm crit} \sim 1$ kG, consistent with
data \citep{2009ApJ...692..538R}, then $\mdot \sim 10^{-7}
M_{\odot}$yr$^{-1}$, which is comparable to values during the initial
disk accretion phase and far too large for stellar winds.  This simply
tells us that, for radial fields, the standard values of $K_V$,
$B_{\rm crit}$ and $\mdot$ yield a $\mathcal{C}$ too large, i.e., too
high a rate of angular momentum loss (because larger $\mathcal{C}$
implies shorter spin-down timescales; see equation [6]).  There are 2
possible resolutions. {\it (1)} The standard values must be modified.
For instance, if $B_{\rm crit} \sim$ 1 kG, as seems likely, then we
may have $K_V \sim 10$ and $\mdot \sim
10^{-10}$\,M$_{\odot}$\,yr$^{-1}$, i.e., Alfv\'{e}n velocities
somewhat higher than escape, and average mass loss rates much higher
than current solar (agreeing with some simulations of $\mdot$ in VLMS
and PMS solar-mass stars; Vidotto et al.\,\,2011 and references
therein).  {\it (2)} Radial fields, which yield the highest rate of
angular momentum loss (since they have the lowest possible multipole
order), are less applicable than higher order multipoles.  Given that
mulipole orders higher than radial are indeed broadly consistent with
field configuration data at all stellar masses (e.g., Donati \&
Landstreet 2009), the latter solution must be important at some level,
regardless of additional variations in $K_V$ and $\mdot$.  We explore
the effect of more complex field geometries in our next paper; further
improvements will doubtless result from more observations as well as
advances in theory and simulations.  Nevertheless, it is heartening
that the very simple theory presented here is able to {\it (a)}
reproduce the broad observational picture of angular momentum
evolution from solar-type stars to VLMS, {\it (b)} reveals the
importance of additional secondary effects such as core-envelope
decoupling and mass-dependent overturn timescales, and {\it (c)} puts
us in a position to quantitatively probe the remaining unknowns, such
as $\mdot$ and field configuration, in the future.

%% In a manner similar to \objectname authors can provide links to dataset
%% hosted at participating data centers via the \dataset{} command.  The
%% second curly bracket argument is printed in the text while the first
%% parentheses argument serves as the valid data set identifier.  Large
%% lists of data set are best provided in a table (see Table 3 for an example).
%% Valid data set identifiers should be obtained from the data center that
%% is currently hosting the data.

\acknowledgements A.R.\ acknowledges financial support from the Deutsche
Forschungsgemeinschaft (DFG) under an Emmy Noether fellowship (RE 1664/4-1)
and a Heisenberg Professorship (RE 1664/9-1). S.M.\ is very grateful to the
{\it International Summer Institute for Modeling in Astrophysics} (ISIMA) for
affording him the time and research environment to complete this work, and
acknowledges the funding support of STFC grant ST/H00307X/1. The authors would
also like to thank the anonymous referee for a detailed and extremely helpful
reading of the paper, which helped to improve it considerably.




%\clearpage

\bibliographystyle{apj}
\bibliography{BrakingApJ}

\clearpage

\end{document}